\documentclass[reprint,prl,aps,amsmath,amssymb,floatfix,superscriptaddress,longbibliography,cite]{revtex4-2}
 
\usepackage{mathtools,amsmath}
\usepackage{graphicx} 
\usepackage{tikz}
\usepackage[breaklinks=true,colorlinks,citecolor=teal,linkcolor=teal,urlcolor=teal]{hyperref}
\usepackage{physics}
\usepackage{siunitx}
\usepackage{bbold}
\usepackage{bibunits}
\usepackage{soul}
\usepackage{bm}
\usepackage[normalem]{ulem}
\usepackage{times}
\usepackage{enumerate}  
\usepackage{float}
\usepackage{amssymb}
\usepackage{xcolor}

\begin{document}

\title{Flux Pumped Kerr-Free Parametric Amplifier}

\author{Kagan Yanik}
\affiliation{Department of Physics and Astronomy, University of Rochester, Rochester, NY 14627, USA}
\affiliation{Institute for Quantum Studies, Chapman University, Orange, CA 92866, USA}

\author{Irwin Huang}
\affiliation{Department of Physics and Astronomy, University of Rochester, Rochester, NY 14627, USA}
\affiliation{Institute for Quantum Studies, Chapman University, Orange, CA 92866, USA}

\author{Bibek Bhandari}
\email{Correspondence should be addressed to:     bbhandari@chapman.edu} 
\affiliation{Institute for Quantum Studies, Chapman University, Orange, CA 92866, USA}

\author{Bingcheng Qing}
\affiliation{Quantum Nanoelectronics Laboratory, Department of Physics, University of California at Berkeley, Berkeley, CA 94720, USA}

\author{Ahmed Hajr}
\affiliation{Quantum Nanoelectronics Laboratory, Department of Physics, University of California at Berkeley, Berkeley, CA 94720, USA}
\affiliation{Graduate Group in Applied Science and Technology, University of California at Berkeley, Berkeley, CA 94720, USA }

\author{Ke Wang}
\affiliation{Quantum Nanoelectronics Laboratory, Department of Physics, University of California at Berkeley, Berkeley, CA 94720, USA}

\author{David I Santiago}
\affiliation{Quantum Nanoelectronics Laboratory, Department of Physics, University of California at Berkeley, Berkeley, CA 94720, USA}
\affiliation{Computational Research Division, Lawrence Berkeley National Laboratory, Berkeley, Berkeley, CA 94720, USA}

\author{Irfan Siddiqi}
\affiliation{Quantum Nanoelectronics Laboratory, Department of Physics, University of California at Berkeley, Berkeley, CA 94720, USA}
\affiliation{Computational Research Division, Lawrence Berkeley National Laboratory, Berkeley, Berkeley, CA 94720, USA}

\author{Justin Dressel}
\affiliation{Schmid College of Science and Technology, Chapman University, Orange, CA 92866, USA}
\affiliation{Institute for Quantum Studies, Chapman University, Orange, CA 92866, USA}

\author{Andrew N Jordan}
\affiliation{The Kennedy Chair in Physics, Chapman University, Orange, CA 92866, USA}
\affiliation{Schmid College of Science and Technology, Chapman University, Orange, CA 92866, USA}
\affiliation{Institute for Quantum Studies, Chapman University, Orange, CA 92866, USA}
\affiliation{Department of Physics and Astronomy, University of Rochester, Rochester, NY 14627, USA}

\begin{abstract}
We propose a flux-pumped superconducting parametric amplifier based on symmetrically threaded superconducting quantum interference devices (SQUIDs) that achieves a Kerr-free operating point under suitable drive conditions. Eliminating the Kerr nonlinearity is advantageous for quantum-limited amplification, as it mitigates unwanted distortions in squeezing and prevents degradation of both gain and quantum efficiency in the high-gain strong drive regime. By replacing the central junction in the symmetrically threaded SQUIDs (STS) configuration with a linear inductor, we find that the Kerr-nonlinearity can be eliminated and the effective Hamiltonian reduces to that of a degenerate parametric amplifier (DPA), up to higher-order corrections in the zero-point fluctuations of the superconducting phase operator. We show that the deviations from ideal DPA behavior introduced by these higher-order terms are significantly weaker than those associated with a Kerr nonlinearity. Consequently, the STS design can be driven strongly while maintaining near-quantum-limited performance at the Kerr-free point. Our analysis predicts phase-preserving gain and efficiency approaching the quantum limit, with robust operation demonstrated up to 25 dB of gain.
\end{abstract}

\maketitle

%\begin{widetext}  

{\em Introduction---} Superconducting qubits have emerged as leading candidates for achieving scalable, fault-tolerant quantum processors \cite{Wallraff2004,PhysRevResearch5L042024,Lu2023,dorogov2022application,PhysRevLett124067701,10186961,10106350127375,devoret2013superconducting,huang2020superconducting,long2024}. The accuracy of quantum algorithms and error correction in these devices, however, rely heavily on fast, high-fidelity single-shot measurement of the qubit states~\cite{dorogov2022application,PhysRevLett124067701,10186961,10106350127375,Krantz2016single}. This has driven research for the development of high-gain and high-efficiency microwave parametric amplifiers for superconducting qubit readouts~\cite{ 9865444,9134828,boutin2017effect, caves1981quantum,caves1982quantum,andrew_book}. These devices rely on wave-mixing processes, where a high-frequency pump interacts with a low-frequency signal via a nonlinear medium, transferring energy to amplify the signal. Parametric amplifiers can operate either in the phase preserving \cite{walls2008quantum,Scully_Zubairy_1997,Liu2024fully, Bergeal2010, roy2018quantum,PhysRevApplied10054020} or phase sensitive mode \cite{malnou2018optimal,Gaikwad2023observing,boutin2017effect,qing2024broadband}. In this letter, we focus on phase-preserving parametric amplifiers which amplify both quadratures symmetrically and can provide phase-independent quantum-limited amplification suitable for multi-qubit readout \cite{10106350127375,PhysRevLett124067701,PhysRevA101042336}.

\begin{figure}[t]
    \centering
    \includegraphics[scale=0.9]{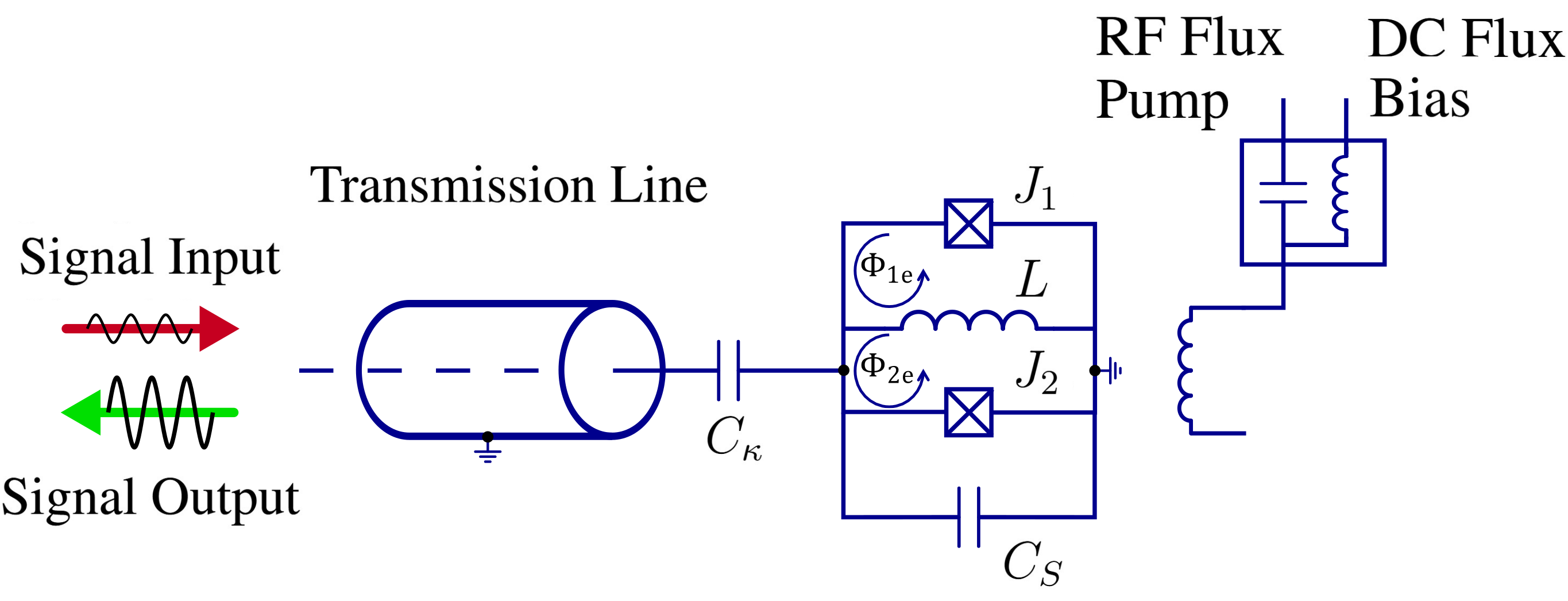}
    \caption{Circuit of a lossless lumped-element flux-pumped Kerr-free STS amplifier coupled to a transmission line propagating the signal input and output fields. The outer branches are composed of junctions ${J}_{1}$ and ${ J}_{2}$ whereas the middle branch consists of an inductor $L$. The double SQUID loop is shunted by a capacitor $C_{\rm S}$ and threaded by an external flux consisting of dc flux bias and ac flux modulation via pump line inductively coupled to the STS. The loop between $J_{1(2)}$ and $L$ is threaded by an external flux $\Phi_{\rm 1(2)e}$.}
    \label{fig:circuit}
\end{figure}

Josephson parametric amplifiers (JPAs) are widely used to achieve near quantum-limited phase preserving amplification in superconducting qubit readout \cite{10106350127375,PhysRevLett106110502,PhysRevA101042336}. JPAs can be operated using two pumping schemes: current pumping and flux pumping. In the current-pumping scheme, the intrinsic nonlinearity of the Josephson junctions is exploited to directly modulate the current through the device, typically by driving the pump at the same frequency as the cavity resonance frequency~\cite{dai2025optimizing,grebel2021flux,roy2015broadband,He2023simulation}. However, current pumping suffers from two major drawbacks. First, the pump and output signal share the same port and same frequency, making it difficult to separate them~\cite{boutin2017effect,kamal2009signal,eichler2013experimental}. Pump leakage into the output path adds unwanted noise beyond the minimum allowed by quantum mechanics, necessitating additional filtering and complicating the path to quantum-limited performance~\cite{boutin2017effect,dai2025optimizing}. Bichromatic current pumping was also explored in Ref.~\cite{boutin2017effect}, where two far detuned current pumps are applied such that their combined frequency equals twice the cavity resonance frequency. This approach provides spectral separation between the pump and signal, allowing the pump tones to be filtered out before measurement. Nevertheless, this scheme still induces a pump-dependent shift of the cavity resonance frequency. Second, the current pump induces a large shift in the cavity resonance frequency~\cite{boutin2017effect} resulting in a higher parametric threshold for maximum gain. In contrast, flux pumping modulates the effective inductance of the JPA by varying the magnetic flux threading the dc superconducting quantum interference devices (SQUID) loop~\cite{boutin2017effect,yamamoto2008flux,xu2025dynamic}. This scheme overcomes the limitations of current pumping since (i) the pump and signal naturally couple to different physical ports, which strongly suppresses pump leakage and reduces added noise~\cite{boutin2017effect,dai2025optimizing,yamamoto2008flux,mutus2014strong}, and (ii) flux-pumping produces a much smaller cavity detuning compared to current pumping~\cite{boutin2017effect}.

In many previous studies, JPAs have been treated as ideal quantum-limited amplifiers, with higher-order nonlinear effects neglected \cite{boutin2017effect}. However, Refs.~\cite{boutin2017effect,kochetov2015higher,10106315003032} demonstrate that these higher-order corrections introduce nonidealities that degrade JPA performance, leading to deviations from the quantum limit in the high-gain regime and imposing constraints on the saturation power. The dominant source of these deviations is the Kerr nonlinearity~\cite{PhysRevApplied11054060}. Near–quantum-limited operation in the Kerr-free regime has been achieved in current-pumped amplifiers based on superconducting nonlinear asymmetric inductive element (SNAIL) \cite{10106350083350,PhysRevApplied11054060,hajr2024,qing2024} and Josephson parametric converter with an inductively shunted Josephson ring modulator (JRMs)~\cite{PhysRevA101042336}. These amplifiers employ a three-wave current pumping scheme which provides a large spectral separation of signal and pump. However, to our knowledge, an analogous Kerr-free implementation based on flux pumping has not yet been demonstrated.  

In this letter, we propose the Symmetrically Threaded SQUIDs (STS) architecture, introduced in Ref.~\cite{bhandari2024symmetrically}, as a flux-pumped platform for realizing a Kerr-free amplifier operating near the quantum limit. In our implementation, the Josephson junction in the central branch of the STS~\cite{bhandari2024symmetrically} is replaced by a linear inductor, which cancels the Kerr nonlinearity of the device for a particular static drive. A comparable design, referred to as linear inductive coupler (LINC) in Ref.~\cite{maiti2025linearquantumcouplerclean}, has been used as a nonlinear coupling element that remains essentially linear when idle and activates its nonlinearity only under drive to implement gates. We first analyze the higher order nonidealities of conventional single-loop flux-pumped dc SQUID amplifiers and establish that Kerr nonlinearity is the dominant nonideality that needs to be mitigated to approach quantum limited amplification. Since Kerr nonideality cannot be completely canceled in single-loop SQUID amplifiers, we propose the STS amplifier design as a solution to operating at Kerr-free regime. We find that the STS amplifier achieves phase-preserving gain and quantum efficiency that closely match those of an ideal degenerate parametric amplifier (DPA), demonstrating near–quantum-limited performance. In addition, our gain and efficiency analysis shows that the Kerr-free STS design surpasses conventional flux-pumped JPAs based on a single-loop dc SQUID, whose operation is constrained by Kerr-induced nonidealities.

{\em Ideal quantum-limited amplifiers---} Ideal quantum-limited microwave amplifiers are often modeled after the degenerate parametric amplifier (DPA), which consists of a single-mode cavity containing a nonlinear medium and is pumped at twice its resonance frequency through a two-photon squeezing drive~\cite{caves1982quantum,walls2008quantum}. In its ideal form, the DPA Hamiltonian includes only the harmonic oscillator term and the two-photon squeezing interaction
\begin{equation}
    \hat H_{\rm DPA} = \Delta \hat a^\dagger \hat a + \frac{\lambda}{2} \hat a^{\dagger^2} + \frac{\lambda^*}{2} \hat a^2,
\end{equation}
where $\hat{a}$ $(\hat{a}^{\dagger})$ denotes the intracavity field annihilation (creation) operator, the detuning $\Delta=\omega_{\rm a}-\omega_{\rm p}/2$, $\omega_{\rm a}$ is the cavity resonance frequency, $\omega_{\rm p}$ is the pump frequency, and $\lambda$ is the two-photon drive strength. Phase-preserving DPAs operate at the quantum limit when cavity dissipation stems solely from the coupling strength of the input signal~\cite{boutin2017effect}. The phase-preserving DPA gain is given by~ \cite{boutin2017effect,roy2015broadband}
\begin{equation}
    G_{\rm DPA}=\left|\frac{\kappa\overline{\kappa}/2-i\kappa(\Delta+\omega)}{\Delta^2(\overline{\kappa}/2-i\omega)^2-|\lambda|^2}-1\right|^2,
    \label{eq:GDPA}
\end{equation}
where $\omega=\omega_{\rm s}-\omega_{\rm p}/2$, $\omega_{\rm s}$ is the signal frequency, $\overline{\kappa}=\kappa+\gamma$ is the total damping rate of the cavity, $\kappa$ is the coupling rate of the signal to the cavity, and $\gamma$ is the coupling rate of additional noises to the cavity addressing the undesired losses~\cite{boutin2017effect,walls2008quantum}. DPA is quantum-limited when there are no undesired losses ($\gamma=0$)~\cite{boutin2017effect}. Moreover, the gain diverges at the parametric threshold $\lambda_{\rm crit}=\sqrt{\Delta^2+\overline{\kappa}^2/4}$~\cite{boutin2017effect,wustmann2013parametric,laflamme2011quantum}. Hence, the amplifier needs to be operated at a value smaller than but near $\lambda_{\rm crit}$ for high gain.

{\em Flux-pumped SQUID-based architectures---} Efforts to realize degenerate parametric amplification in superconducting architectures have employed SQUIDs, which may be operated as charge-pumped or flux-pumped devices depending on how they are modulated. The most basic flux-pumped superconducting device consists of a single-loop dc SQUID~\cite{zorin2019flux,yamamoto2008flux,planat2019understanding,hatridge2011dispersive}. An external magnetic flux threads the SQUID loop and serves as the pump drive at frequency $\omega_{\rm p}$. This pump participates in a degenerate three-wave mixing process involving the signal and idler modes at frequencies $\omega_{\rm s}$ and $\omega_{\rm i}$, respectively, satisfying $\omega_{\rm p}=\omega_{\rm s}+\omega_{\rm i}\approx 2\omega_{\rm s}$. When the bare cavity resonance is tuned near $\omega_{\rm a}\approx \omega_{\rm s}$, the flux modulation generates an effective two-photon drive. Operationally, this is achieved by biasing the SQUID at a static flux $F$ and applying a small AC modulation such that the external flux takes the form $\Phi_{\rm e}/\varphi_0 = F + \delta f \cos(\omega_{\rm p} t)$, where $\delta f$ is the modulation depth~\cite{boutin2017effect} and $\varphi_0 = h/2e$ is the flux quantum. A fundamental limitation of the single-loop SQUID design is that its Kerr nonlinearity cannot be mitigated~\cite{bhandari2024symmetrically,hua2025engineering}. 

To enable Kerr cancellation, one must move beyond the minimal single-loop design. Multi-loop SQUID geometries introduce additional tunability that allows the leading-order nonlinearities to be engineered away. In such flux-pumped multi-loop SQUIDs, the static effective Hamiltonian expanded to fourth order in the zero-point phase fluctuations takes the general form~\cite{hua2025engineering}
\begin{equation}
 \hat H_{\rm SQ} = \hat H_{\rm DPA}+ K \hat a^{\dagger^2} \hat a^2 +\Lambda \left(\hat a^{\dagger^3} \hat a +\hat a^{\dagger} \hat a^3 \right)+\zeta\left(\hat{a}^{\dagger^4}+\hat{a}^4\right),
 \label{eq:HSQ}
 \end{equation}
where $K$ is the Kerr-nonlinearity, and $\Lambda$ and $\zeta$ capture the other flux-dependent nonlinear corrections inherited from the underlying SQUID potential. A particularly useful realization is the symmetric two-loop SQUID (STS) circuit containing a linear inductor between the loops, as illustrated in Fig.~\ref{fig:circuit}. When both loops of STS are threaded by the same external flux modulation,
$\Phi_{\rm 1e}/\varphi_0 = \Phi_{\rm 2e}/\varphi_0 = F + \delta f \cos(\omega_{\rm p} t)$,
the Kerr coefficient becomes flux dependent and takes the form $K = -E_{\rm C} \cos F / 2$, where $E_{\rm C}$ is the charging energy of the circuit. Hence, the Kerr-coefficient vanishes at $F=-\pi/2$. At this same bias point, the quartic correction also cancels because $\zeta \propto \cos F$, leaving only the cubic nonlinear term (see the Supplemental Material for details). Consequently, the static effective Hamiltonian of the STS device simplifies to
\begin{equation}
    \hat H_{\rm STS} = \hat H_{\rm DPA}+\Lambda \left(\hat a^{\dagger^3} \hat a +\hat a^{\dagger} \hat a^3 \right),
 \label{eq:H_STS}
\end{equation}
representing a substantial reduction in unwanted nonlinearities. However, since $\Lambda \propto \delta f \sin F$ (see Supplemental Material) takes a maximum at $F=\pi/2$, we will study in detail its impact on parametric amplification. It is important to note that a conventional single-loop SQUID cannot be biased at the symmetry point $F=-\pi/2$ because this corresponds to its ``off" point where the Josephson energy is minimized and the resonance frequency approaches zero. As a result, single-loop flux-pumped JPA inevitably retains all nonlinear corrections in Eq.~(\ref{eq:HSQ}), including a nonzero Kerr nonlinearity,  while the STS design permits systematic suppression of these terms. Throughout the remainder of this work, we refer to single-loop SQUID–based parametric amplifiers as JPA, and to double-loop STS–based devices as STS. The explicit expressions for $K$, $\Lambda$, and $\zeta$ for both circuit designs are provided in the Supplemental Material.

{\em Nonidealities in JPA---}
Leading higher-order nonlinearities characterized by the coefficients $K$ and $\Lambda$ lead JPAs to deviate from ideal DPA dynamics, reducing their achievable squeezing and overall performance. This deviation is captured by comparing the JPA’s moments to those of an ideal DPA through the parameter~\cite{boutin2017effect}
\begin{equation}
    \Xi_{i}=1-\frac{|M_{i}|}{\sqrt{N_{i}(N_{i}+1/2)}},
\end{equation}
where $i\in \{{\rm DPA,~JPA}\}$, $M_{\rm i}=\langle\hat{a}^2\rangle-\langle\hat{a}\rangle^2$ and $N_{\rm i}=\langle\hat{a}^{\dagger}\hat{a}\rangle-|\langle\hat{a}\rangle|^2$, $\Xi_{\rm DPA} =0$ and $\Xi_{\rm JPA}\in[0,1]$. By construction, $\Xi_{\rm JPA}=0$ corresponds to ideal DPA-level performance, while $\Xi_{\rm JPA}\to 1$ indicates increasingly strong deviation from DPA behavior. Since we work in the weak–modulation regime ($\delta f \ll 1$) and the quartic term in $\hat H_{\rm SQ}$ scales as $\zeta \propto \delta f^{2}$ (whereas $K\propto \delta f^0$ and $\Lambda \propto \delta f^1$)
, we neglect the contribution from $\zeta$ in the remainder of this letter. To understand the limitations of single-loop SQUID JPAs and motivate the need for a double-loop STS architecture, we first examine the deviation parameter $\Xi_{\rm JPA}$ for various nonlinear contributions. 

\begin{figure}[t]
    \centering
    \includegraphics[width=\linewidth,scale=1.2]{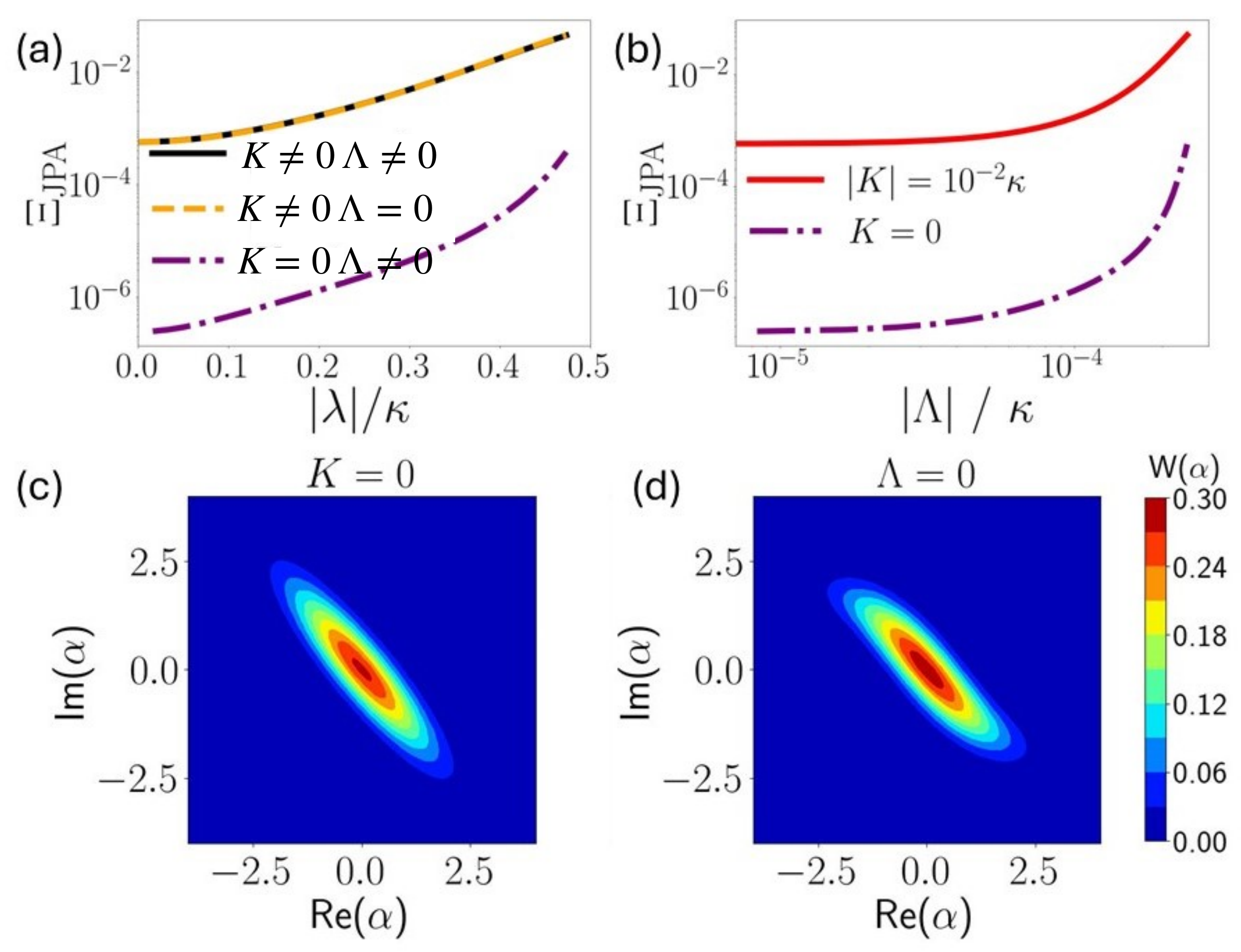}
    \caption{(a) Deviation of JPAs from ideal DPA dynamics $(\Xi_{\rm JPA})$ plotted as a function of drive strength $\lambda$. For $K\neq 0$, we consider $|K|/\kappa=10^{-2}$ and $|\lambda|/\kappa\in[0,0.47]$ corresponds to $|\Lambda|/\kappa\in[0,2.36\times10^{-4}]$. (b) Deviation of JPAs plotted for $|K|=10^{-2}\kappa$, and $K=0$ as a function of $\Lambda$. Steady-state Wigner function plot for $|\lambda|=0.45\kappa$ (indicated by stars in (a)) in the (c) Kerr-free case, $K=0$, and (d) for $\Lambda = 0$. For (c), we consider $|\Lambda|/\kappa=2.25\times10^{-4}$, whereas for (d), we consider $|K|/\kappa=10^{-2}$. Note that, the $K=0$ case can only be realized with an STS circuit. We take $\Delta=0$ for all subfigures.}
    \label{fig:Dev_DPA}
\end{figure}

Figure~\ref{fig:Dev_DPA}(a) shows the deviation parameter $\Xi_{\rm JPA}$, computed from the Hamiltonian in Eq.~(\ref{eq:HSQ}), as a function of the two-photon drive strength $\lambda$. We consider three scenarios: (i) both the Kerr and $\Lambda$ nonlinearities present (black solid curve), (ii) only the Kerr nonlinearity included (orange dashed curve), and (iii) only the $\Lambda$ correction retained as in the STS Hamiltonian (purple dot-dashed curve). Note that, The near-perfect overlap of the black and orange curves demonstrates that $\Lambda$ has a negligible effect on $\Xi_{\rm JPA}$ over the parameter range studied, $|\lambda|/\kappa \in [0,\,0.47]$. Note that, for $\Delta = 0$, the gain in Eq.~(\ref{eq:GDPA}) diverges at $|\lambda_{\rm crit}|  = 0.5\kappa$. The upper bound on $\lambda$ is chosen to ensure at least $25\,\mathrm{dB}$ of gain for the degenerate parametric amplifier (DPA). Since $\Lambda = (\Phi_{\rm zps}/\varphi_0)^2 \lambda/6$ (see the Supplementary Material), where the zero point phase spread, $\Phi_{\rm zps}/\varphi_0\ll 1$, this corresponds to $|\Lambda|/\kappa \in [0,\,2.36\times10^{-4}]$ for the case $\Lambda \neq 0$. By contrast, eliminating the Kerr nonlinearity, as realized in the STS Hamiltonian, leads to a substantial reduction in $\Xi_{\rm JPA}$, bringing the system much closer to ideal DPA behavior. In Fig.~\ref{fig:Dev_DPA}(b), we examine $\Xi_{\rm JPA}$ as a function of $\Lambda$ for different values of the Kerr coefficient. We find that $\Xi_{\rm JPA}$ is significantly larger in the presence of Kerr nonlinearity (red solid curve) than in its absence (purple dot-dashed curve), particularly in the weak two-photon driving regime. Nevertheless, even when $K=0$, $\Xi_{\rm JPA}$ increases sharply for $\Lambda/\kappa \gtrsim 10^{-4}$ (as illustrated by the rise of purple dot-dashed curve). However, we will later show that deviations induced by $\Lambda$ have a weaker effect on the amplifier gain and efficiency than those arising from Kerr nonlinearity. These observations are further corroborated by the Wigner-function plots in Fig.~\ref{fig:Dev_DPA}(c) and (d). Removing the Kerr term, as realized in the STS Hamiltonian, yields a nearly ideal squeezed state (Fig.~\ref{fig:Dev_DPA}(c)), whereas suppressing $\Lambda$ alone while retaining Kerr results in strong state distortion (Fig.~\ref{fig:Dev_DPA}(d)). Thus, we conclude that the Kerr term in the Hamiltonian leads to stronger deviations than the cubic $\Lambda$ term to the DPA Hamiltonian.

\begin{figure}[t]
    \centering
    \includegraphics[width=\linewidth]{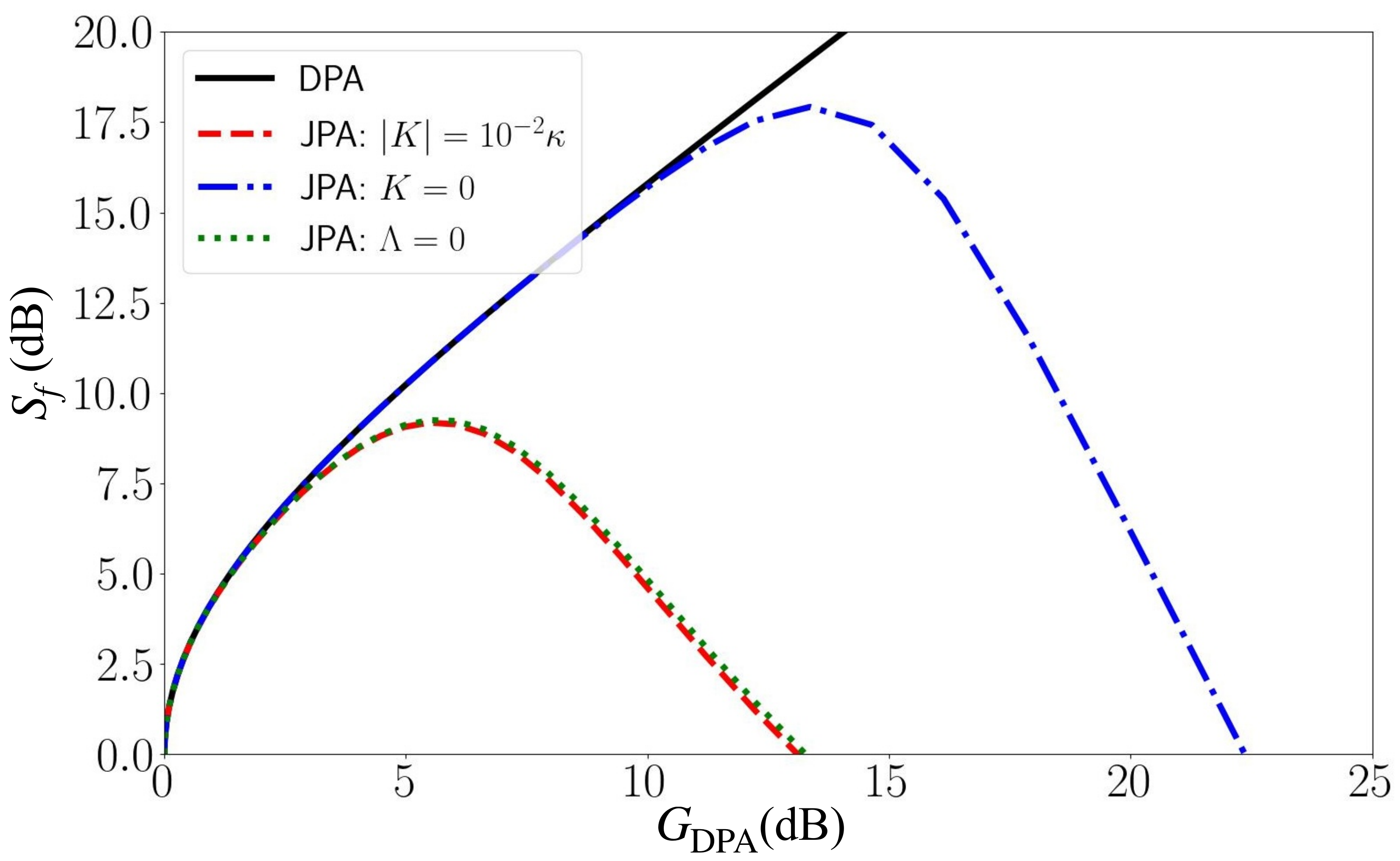}
    \caption{Squeezing level for DPA, and JPAs plotted as a function of phase-preserving DPA gain. We consider, $\Delta=\gamma=0$, $\kappa/2\pi=300\space{\rm MHz}$, Josephson inductance $L_{J}=80\space{\rm pH}$, and capacitance $C_{\Sigma}=2\space{\rm pF}$.}
    \label{fig:Sq_level}
\end{figure}

To further quantify how circuit nonidealities drive deviations from the DPA limit, we evaluate the squeezing level
$S_f={\langle\Delta\hat{X}_{\rm vac}^2\rangle}/{\langle\Delta\hat{X}_{\rm min}^2\rangle}$,
where $\langle\Delta\hat{X}_{\rm vac}^2\rangle = 1/2$ is the variance of the vacuum state and $\langle\Delta\hat{X}_{\rm min}^2\rangle$ denotes the minimum variance of the output-field X quadrature~\cite{boutin2017effect,Zhong2013squeezing}. This metric provides a quantitative measure of how strongly each nonideality degrades squeezing. Figure~\ref{fig:Sq_level} shows $S_f$ as a function of DPA gain for four cases: an ideal DPA (black solid curve), a JPA including both nonidealities $K$ and $\Lambda$ (red dashed curve), a JPA with $K=0$ as in the STS case (blue dot-dashed curve), and a JPA with $\Lambda=0$ (green dotted curve). We observe that when both K and $\Lambda$ are present, the squeezing level deviates markedly from the DPA benchmark beyond $\sim 5 $dB of gain, falling much below the ideal DPA squeezing. Eliminating $\Lambda$ alone while retaining $K$ (green dotted curve) yields nearly identical behavior, confirming that Kerr nonlinearity is the dominant source of squeezing degradation in this regime. In contrast, a Kerr-free JPA (blue dot-dashed curve) closely follows the DPA squeezing level up to approximately 15 dB of gain and exhibits a substantial improvement over the other configurations. Together, these results underscore the central role of Kerr nonlinearity in limiting the performance of single-SQUID JPAs and motivate the adoption of multi-loop circuit designs that enable systematic Kerr suppression.

{\em Kerr-free STS parametric amplifier---} The quantum Langevin equation for the STS parametric amplifier, as shown in Fig.~(\ref{fig:circuit}), can be written as $\dot{\hat{a}}=i[\hat{H}_{\rm STS},\hat{a}]-{\Bar{\kappa}}/{2}\hat{a}+\sqrt{\kappa}\hat{a}_{\rm in}+\sqrt{\gamma}\hat{b}_{\rm in}$, where $\hat a_{\rm in}$ and $\hat b_{\rm in}$ are the signal and undesired loss input fields, respectively. Since we compare the performance of the STS to a quantum-limited DPA ($\gamma=0$), we assume a lossless cavity ($\gamma=0$) for all amplifiers discussed in this letter (see Supplemental Material for analysis of the STS amplifier with $\gamma\neq 0$). The corresponding input–output relation is given by $\hat a_{\rm out}(t)=\sqrt{\kappa}\hat a(t) - \hat a_{\rm in}(t)$, which defines the output field carrying the amplified signal, $\hat a_{\rm out}$.

\begin{figure}[h]
    \centering
    \includegraphics[width=0.9\linewidth]{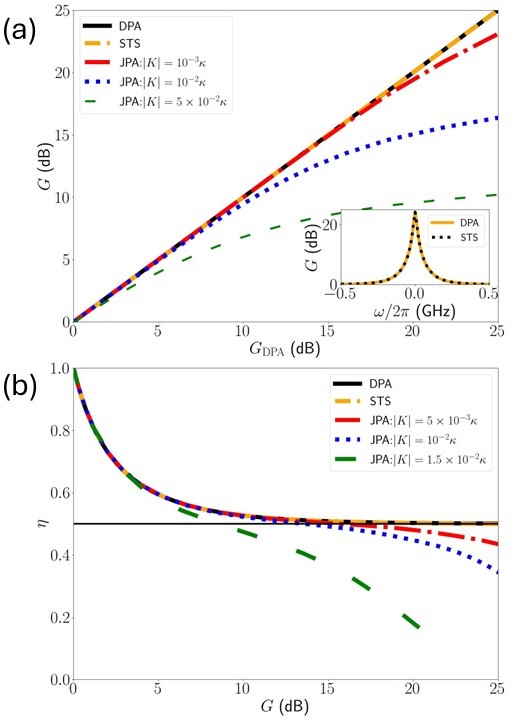}
    \caption{(a) Phase-preserving gain plotted against DPA gain and (b) Phase-preserving quantum efficiency plotted at $\omega=0$ against the given design's gain for DPA, Kerr-free STS amplifier, and single-SQUID JPAs with varying Kerr nonlinearities. Inset in (a) shows the phase-preserving gain calculated using Eq.~(\ref{eq:gain analytical}) against signal detuning $\omega$ for DPA and STS amplifier. We take $\Delta=\gamma=0,\kappa/2\pi=300 \space {\rm MHz}$, and the sweep for the two-photon drive $0\leq\lambda<\kappa/2$. For STS, we consider capacitance $C_\Sigma=4\space {\rm pF}$, Josephson inductance $L_{\rm J}=80\space{\rm pH}$, and linear inductance $L=100\space{\rm pH}$.}
    \label{fig:efficiency}
    \label{fig:ppGcomp}
\end{figure}

Using semiclassical harmonic balance equations and input-output theory \cite{PhysRevApplied10054020}, the phase-preserving gain $G_{\rm STS}(\omega)$ of the STS amplifier as a linear response to the signal probe at frequency $\omega_{\rm s}$ can be expressed as (see the Supplemental Material for harmonic balance and stability diagram analysis)
\begin{equation}
   \small{ G_{\rm STS}(\omega)=\left|\frac{i\kappa\left(\omega+\Delta_{\rm eff}^*+i\frac{\kappa}{2}\right)}{\left(\omega-\Delta_{\rm eff}+i\frac{\kappa}{2}\right)\left(\omega+\Delta_{\rm eff}^*+i\frac{\kappa}{2}\right)+|\lambda_{\rm eff}|^2}-1\right|^2.}
    \label{eq:gain analytical}
\end{equation}
Here, $\omega = \omega_{\rm s} - \omega_{\rm p}/2$ denotes the detuning of the signal from the rotating-frame frequency, while $\Delta_{\rm eff} = \Delta + 3\Lambda \alpha_{\rm i}\alpha_{\rm s}$ and $\lambda_{\rm eff} = \lambda + 9\Lambda n_{\rm s}$ represent the effective detuning and two-photon drive strength, respectively. The mean intracavity field amplitudes at the signal and idler frequencies $\omega_{\rm s,i}$ are denoted by $\alpha_{\rm s,i}$, with the corresponding photon numbers given by $n_{\rm s(i)} = |\alpha_{\rm s(i)}|^2$. We find that the gain expression of the STS amplifier has the same functional form as that of an ideal DPA in Eq.~(\ref{eq:GDPA}), with the only difference being the presence of $\Lambda$-induced corrections to the effective detuning and drive strength. When the effective STS parameters are matched to those of the DPA ($\lambda_{\rm eff} = \lambda_{\rm DPA}$, $\Delta_{\rm eff} = \Delta_{\rm DPA}$), the STS amplifier exhibits the same gain–bandwidth performance as a quantum-limited DPA, as shown in the inset of Figure.~\ref{fig:ppGcomp}(a).

{\em Gain and efficiency in the linear response regime---}
We study the STS amplifier gain at the center frequency $\omega=\omega_{\rm s}-\omega_{\rm p}/2=0$. To compute the phase-preserving gain of the STS amplifier, we evaluate the cavity-field displacement induced by a narrow-band signal probe and the corresponding output response~\cite{boutin2017effect}. The probe is introduced via the Hamiltonian $\hat H_{\rm probe}=\epsilon_{\rm probe}\hat{a}^{\dagger}+\epsilon_{\rm probe}^{*}\hat{a}$, where $\epsilon_{\rm probe}$ is the probe amplitude. We operate in the linear-response regime by assuming a weak probe, $\epsilon_{\rm probe}\ll\kappa$. Following Ref.~\cite{boutin2017effect}, the input–output relation is then expressed in terms of the phase-sensitive gain matrix as~\cite{boutin2017effect}
\begin{gather}
 \begin{bmatrix} \hat{X}_{\rm out} \\ \hat{P}_{\rm out} \end{bmatrix}
 =
  \begin{bmatrix}
   g_{11} &
   g_{12} \\
   g_{21} &
   g_{22} 
   \end{bmatrix}
   \begin{bmatrix} \hat{X}_{\rm in} \\ \hat{P}_{\rm in} \end{bmatrix},\label{g_matrix}
\end{gather}
where the quadratures are defined as $\hat{X}=(\hat{a}+\hat{a}^{\dagger})/\sqrt{2}$ and $\hat{P}=i(\hat{a}^{\dagger}-\hat{a})/\sqrt{2}$. The phase-preserving gain can then be expressed as~\cite{boutin2017effect},
 $G=|g_{11}+g_{22}+i(g_{21}-g_{12})|^2/4$.

We compute the gain by solving for the steady-state expectation values of the output quadratures with the input field amplitude, $\langle\hat a_{\rm in}\rangle=i\epsilon_{\rm probe}/\sqrt{\kappa}$. The steady state follows from the master equation~\cite{boutin2017effect},
$\dot{\hat \rho}=-i[\hat{H}_{\rm STS}+\hat{H}_{\rm probe},\hat{a}]+\kappa\mathcal{D}[\hat{a}]\hat{\rho}$,
where $\mathcal{D}[\hat{a}]\hat{\rho}=\hat{a}\hat{\rho}\hat{a}^{\dagger}-(\hat{a}^{\dagger}\hat{a}\hat{\rho}+\hat{\rho}\hat{a}^{\dagger}\hat{a})/2$ is the dissipation superoperator.

Figure~\ref{fig:ppGcomp}(a) presents the phase-preserving gain of the STS amplifier alongside that of an ideal DPA and JPAs with varying Kerr nonlinearities. At low gain, the almost ideal JPA (red dot-dashed curve) closely follows the DPA response (solid black curve) and remains quantum-limited up to approximately 15 dB. At higher gain and stronger Kerr nonlinearity, however, the JPA gain falls below the DPA value for the same drive strength $\lambda$, indicating that increasing Kerr nonlinearity restricts the range of quantum-limited operation (compare the dot-dashed red curve with dotted blue and dashed green curves). In contrast, the Kerr-free STS amplifier (dashed orange curve) remains quantum-limited up to 25 dB of gain, demonstrating that the residual nonideality $\Lambda$ is too weak to appreciably degrade its performance.

The phase-preserving quantum efficiency provides a direct measure of an amplifier’s proximity to quantum-limited operation. For a linear quantum amplifier, the ideal efficiency is given by~\cite{caves2012quantum,boutin2017effect,peng2022floquet}
\begin{equation}
\eta=\left(1+2\mathcal{A}\right)^{-1},
\end{equation}
where the added noise number $\mathcal{A}=\langle|\Delta\hat{a}_{\rm in}|^2\rangle/G$ characterizes the mean-square fluctuation of the input signal normalized by the gain~\cite{caves2012quantum}. The fluctuation is defined as $\langle|\Delta\hat{a}|^2\rangle=\langle|{\hat{a},\hat{a}^{\dagger}}|\rangle/2-|\langle\hat{a}\rangle|^2$. The added noise is fundamentally constrained by the Heisenberg uncertainty principle~\cite{caves1981quantum,caves1982quantum,caves2012quantum}, with a lower bound $\mathcal{A}\geq(G-1)/(2G)$, corresponding to half a photon of noise in the high-gain limit.

As shown in Fig.~\ref{fig:efficiency}(b), the DPA (solid black curve) exhibits unit efficiency at zero gain, reflecting the absence of added noise, and converges to an efficiency of 0.5 at gains exceeding 15dB. JPAs with finite Kerr nonlinearity coincide with the DPA in the low-gain regime but require increased added noise to reach higher gains, leading to efficiencies below the quantum limit. This degradation becomes more pronounced with increasing Kerr nonlinearity. In contrast, the Kerr-free STS amplifier (orange dashed curve) remains overlapped with the DPA even in the high-gain regime despite the presence of the $\Lambda$ term. Together, the gain and efficiency results demonstrate that the residual nonideality $\Lambda$ in the STS Hamiltonian is practically negligible over the gain range considered.

We find that the dip appearing in the squeezing level $S_f$ at high gain does not manifest in the gain and efficiency, since these quantities probe different aspects of the output field and exhibit distinct sensitivities to the $\Lambda$ contribution (compare the blue dot-dashed curve in Fig.~\ref{fig:Sq_level} and orange dashed curves in Fig.~\ref{fig:ppGcomp} (a) and (b)). The squeezing level depends exclusively on the variance of the squeezed $\hat X$ quadrature and is therefore highly sensitive to the off-diagonal gain element that arises from quadrature mixing and non-Gaussian distortions. At high gain, where the diagonal element associated with squeezing becomes small, even weak $\Lambda$ can significantly enhance these off-diagonal contributions, leading to a reduction in $S_f$. By contrast, the phase-preserving gain and efficiency are governed primarily by the amplification of the anti-squeezed $\hat P$ quadrature at high gain, which remains the dominant contribution even in the presence of $\Lambda$, thereby mitigating the influence of off-diagonal gain elements to a larger extent. In addition to gain and efficiency performances, we also characterize the stability diagram for the STS amplifier in the space of parameters $\Delta$ and $\lambda$ (see the Supplemental Material). We see that at $G=25$ dB ($\Delta=0$, $|\lambda|/\kappa\approx0.47$), the STS amplifier is in bi-stable region with two stable points. However, since the two stable points are separated by a distance beyond the intracavity population of the amplifier, STS is effectively in a mono-stable region with a trivial ground state, just like a DPA.

In conclusion, using the deviation parameter $\Xi_{\rm JPA}$, intracavity Wigner-function plots, and squeezing level, we have demonstrated that Kerr nonlinearity is the dominant nonideality in single-SQUID JPAs and is primarily responsible for their departure from ideal DPA behavior. We further showed that our flux-pumped STS architecture enables systematic cancellation of the Kerr nonlinearity by operating at a static flux bias $F=-\pi/2$, a regime inaccessible to single-loop SQUID JPAs. Through both analytical expressions and numerical simulations of phase-preserving gain and efficiency, we established that the STS amplifier closely emulates ideal DPA performance and substantially outperforms single-loop SQUID JPAs up to gains of 25 dB. The resulting enhancements in squeezing, gain, and efficiency open new opportunities for high-fidelity superconducting qubit readout and quantum feedback applications~\cite{murch2013observing,vijay2012stabilizing}.

\section{Acknowledgements}
We acknowledge helpful discussions with Kevin P. O'Brien. This work was supported
by the Army Research Office through Grant No. W911NF22-1-0258.

\bibliography{refs_sup}

\clearpage
\pagestyle{empty}
\onecolumngrid
\vspace*{10pt}
\renewcommand{\theequation}{S\arabic{equation}}
\setcounter{equation}{0}
\begin{bibunit}[apsrev4-2]
\begin{center}
	\large \textbf{Supplemental Material for ``Flux Pumped Kerr-Free Parametric Amplifier"}
\end{center}

\section{Flux-pumped Josephson Parametric Amplifiers: Circuit Quantum Electrodynamics}
\label{app:circquant}
In this section, we will study the circuit quantization in the presence of time-dependent flux for a DC SQUID, STS with an inductor in the middle branch as shown in Fig.~\ref{fig:STSwithL}, and STS with a Josephson junction in the middle branch~\cite{bhandari2024symmetrically}. We will show how the STS design with an inductor in the middle branch cancels the Kerr-nonlinearity through proper tuning of the external flux through the SQUID loops. We will also show that a Kerr-free operational point cannot be realized with the other designs considered.

\subsection{DC SQUID}
\label{app:SQUIDquant}
\begin{figure}[h]
    \centering
    \includegraphics[width=0.2\linewidth]{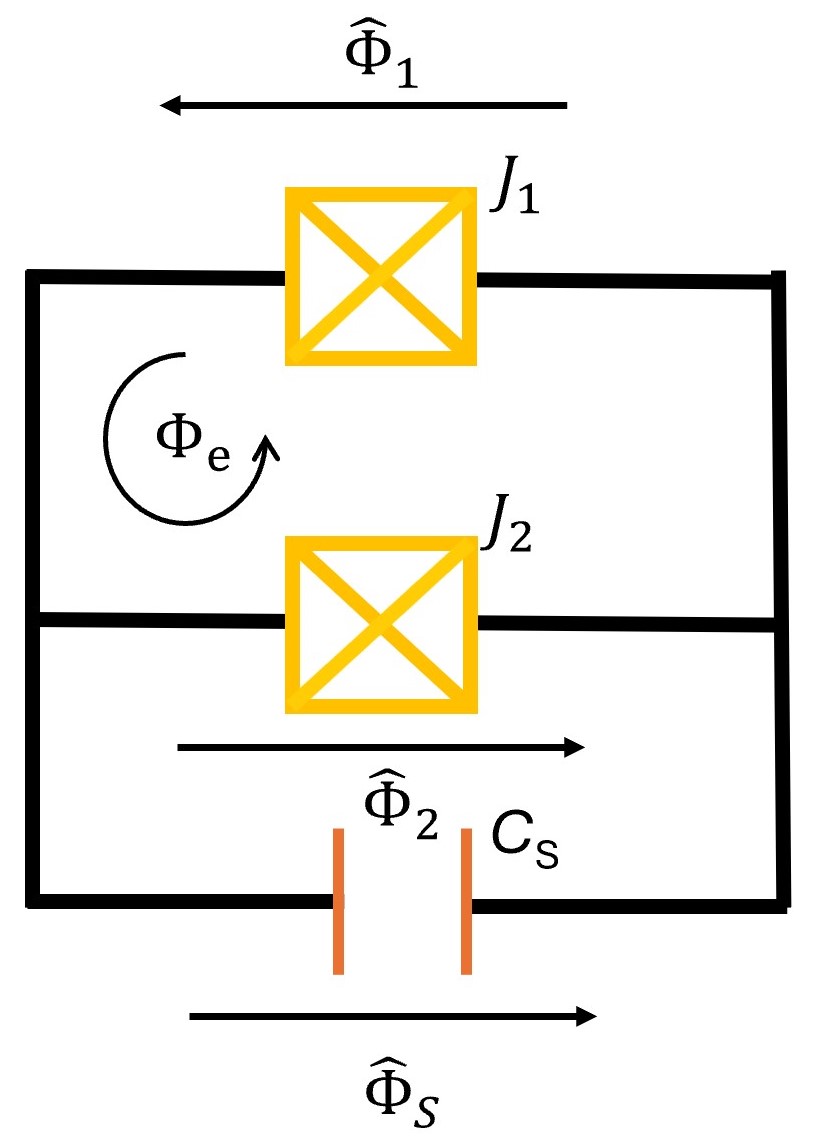}
    \caption{Circuit design of the flux-pumped JPA. The junctions ${\rm J}_{1}$ and ${\rm J}_{2}$ compose the DC SQUID, which is shunted by capacitance $C_{\rm S}$. The loop between ${\rm J}_{1}$ and ${\rm J}_{2}$ is threaded by an external flux $\Phi_{\rm e}$. $\hat{\Phi}_{\rm 1,2,S}$ are the branch flux operators.}
    \label{fig:DC SQUID}
\end{figure}
We consider a flux-pumped Josephson parametric amplifier (JPA) \cite{boutin2017effect}, consisting of a DC SQUID driven by an external magnetic flux $\Phi_{\rm e}$, as illustrated in Fig.~\ref{fig:DC SQUID}. The time-dependent external flux serves as the pumping mechanism essential for the operation of the parametric amplifier.

Following Ref.~\cite{you2019circuit}, the Hamiltonian for a parametrically driven DC SQUID in the lab frame can be written as
\begin{equation}
    \hat{H}_{\rm lab,JPA}=4E_{\rm C}\hat{n}^2-E_{\rm J1}{\rm cos}\left(\frac{\hat{\Phi}-\Phi_{\rm e}/2}{\varphi_0}\right)-E_{\rm J2}{\rm cos}\left(\frac{\hat{\Phi}+\Phi_{\rm e}/2}{\varphi_0}\right),
    \label{eq:ham_JPA}
\end{equation}
where $\varphi_0=\hbar/2e$ is the reduced flux quantum, and the reduced charge operator $\hat{n}=in_0(\hat{a}^{\dagger}-\hat{a})$ and flux operator $\hat{\Phi}=\Phi_{\rm zps}(\hat{a}^{\dagger}+\hat{a})$ are expressed in terms of the bosonic creation $(\hat a^\dagger)$ and annihilation operators $(\hat a)$. Further, $\Phi_{\rm zps}=2\varphi_0\sqrt{E_{\rm C}/\omega_{\rm a}}$ is the zero point flux fluctuations where $\omega_{\rm a}$ is the bare resonator frequency. The external flux, $\Phi_{\rm e}/2\varphi_0=F+\delta f{\rm cos}(\omega_{\rm p}t)$, where $F$ is the static flux, $\omega_{\rm p}\approx2\omega_{\rm a}$ is the pump frequency and $\delta f$ is the pump modulation depth.  $E_{\rm J1}$ and $E_{\rm J2}$ are the junction energies while $C_{\rm 1}$ and $C_{\rm 2}$ are the capacitances of junctions ${\rm J}_1$ and ${\rm J}_2$. The DC SQUID is shunted by capacitance $C_{\rm S}$. The charging energy is defined as $E_{\rm C}= e^2/2C_{\Sigma}$ with $C_\Sigma = C_1 + C_2 + C_{\rm S}$. While deriving Eq.~(\ref{eq:ham_JPA}), we considered $C_1 = C_2$. Assuming symmetric Josephson junctions, $E_{\rm J1}=E_{\rm J2}=E_{\rm J}$, the Hamiltonian reduces to
\begin{equation}
    \hat{H}_{\rm lab,JPA}=4E_{\rm C}\hat{n}^2-2E_{\rm J} \cos\left(\frac{\Phi_{\rm e}}{2\varphi_0}\right){\rm cos}\left(\frac{\hat{\Phi}}{\varphi_0}\right).
    \label{eq:SQUIDbeforeRWA}
\end{equation}
Following the methods in Ref.~\cite{boutin2017effect}, we expand the JPA Hamiltonian up to fourth order in $\Phi_{\rm zps}$ and assume weak modulation depth, $\delta f\approx(\Phi_{\rm zps}/\varphi_0)$. Then evaluating the Hamiltonian in rotating frame at frequency $\omega_{\rm p}/2$ and applying rotating wave approximation (RWA), we obtain~\cite{boutin2017effect}
\begin{equation}
    \hat H_{\rm JPA}=\Delta_{\rm f}\hat{a}^{\dagger}\hat{a}+\frac{\lambda_{\rm f}}{2}(\hat{a}^{\dagger2}+\hat{a}^2)+K_{\rm f}\hat{a}^{\dagger2}\hat{a}^2+\Lambda_{\rm f}(\hat{a}^{\dagger}\hat{a}^3+\hat{a}^{\dagger3}\hat{a})+\zeta_{\rm f}(\hat{a}^{\dagger^4}+\hat{a}^4),
\end{equation}
where $\Delta_{\rm f}=\omega_{\rm a}-\omega_{\rm p}/2+2K_{\rm f}$ is the effective detuning, $\lambda_{\rm f}=E_{\rm J}^{(1)}(\Phi_{\rm zps}/\varphi_0)^2/2$ is the effective pump strength, $K_{\rm f}=-E_{\rm J}^{(0)}(\Phi_{\rm zps}/\varphi_0)^4/4$ is the Kerr nonlinearity, and $\Lambda_{\rm f}=-E_{\rm J}^{(1)}(\Phi_{\rm zps}/\varphi_0)^4/12$ and $\zeta_{\rm f}=-E_{\rm J}^{(2)}(\Phi_{\rm zps}/{\varphi_0})^4/48$ are the extra nonidealities. $E_{\rm J}^{(0)},E_{\rm J}^{(2)}\propto\cos F$, and $E_{\rm J}^{(1)}\propto\sin F$~\cite{boutin2017effect}. Given that Kerr is the leading nonideality that needs to be cancelled to approach quantum-limited amplification~\cite{boutin2017effect}, we need to operate SQUID at static flux $F=-\pi/2$. However, at $\Phi_{\rm e}/2\varphi_0 = \pi/2$, $\hat H_{\rm lab, JPA} = 4E_{\rm C} \hat n^2$, and the SQUID reduces to a capacitor. Hence, the SQUID cannot be operated around $F = \pi/2$ where the Kerr-nonlinearity vanishes.

\subsection{STS Design with an Inductor}
\label{app:STSwithLquant}
\begin{figure}[h]
    \centering
    \includegraphics[width=0.2\linewidth]{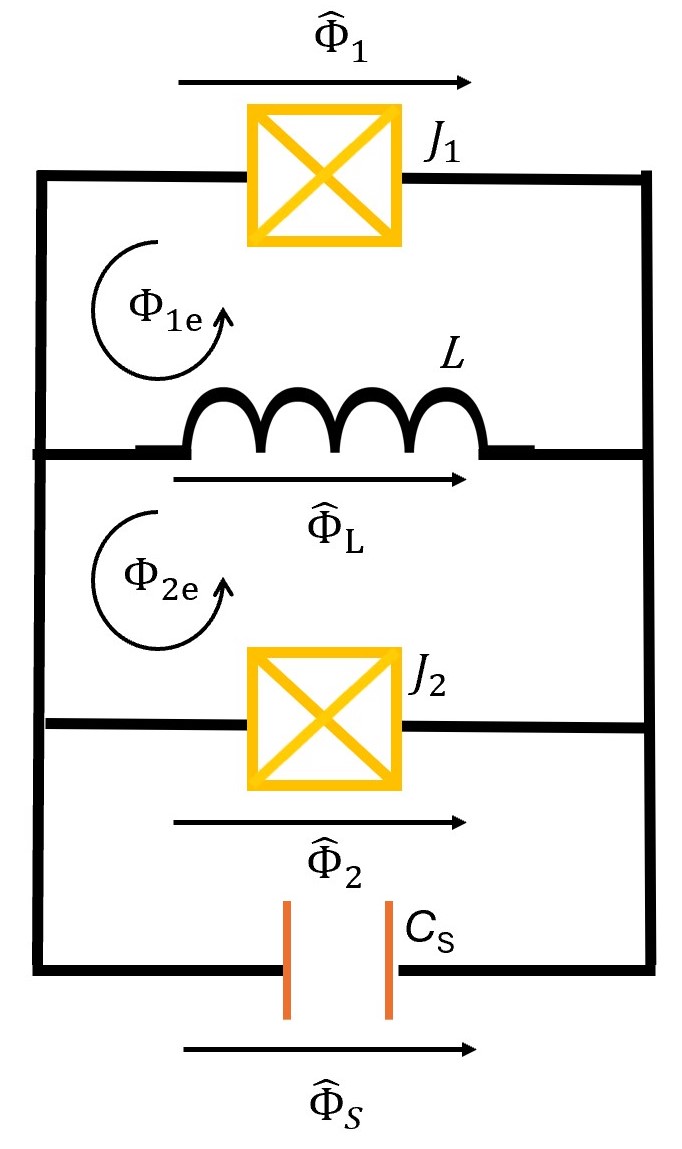}
    \caption{STS circuit with an inductor in the middle branch: The outer branches are composed of junctions ${\rm J}_{1}$ and ${\rm J}_{2}$ whereas the middle branch consists of an inductor L. The double SQUID loop (${\rm J_1 - L - J_2 }$) is shunted by a capacitor $C_{\rm S}$. The loop between ${\rm J}_{1(2)}$ and L is threaded by an external flux $\Phi_{\rm 1(2)e}$. $\hat{\Phi}_{j}$, where $j=\text{1,2,L,S}$, are the branch flux operators.}
    \label{fig:STSwithL}
\end{figure}
In search for a circuit design that can provide a Kerr-free amplifier, we propose a symmetrically threaded SQUID (STS) design discussed in Ref.~\cite{bhandari2024symmetrically}. However, in this STS design, we replaced the Josephson junction in the middle branch with a linear inductor with inductance $L$, as shown in Fig.~\ref{fig:STSwithL}. The bottom branch contains a Josephson junction $\rm J_2$ shunted by capacitance $C_2$ whereas the top branch has a junction $\rm J_1$ and capacitance $C_1$. The circuit is shunted by a capacitance $C_{\rm S}$. Similar to the case of DC SQUID, we start the circuit quantization by expressing the Lagrangian in terms of the branch flux operators. We obtain
\begin{equation}
    \hat{\mathcal{L}}=\frac{1}{2}C_{\rm S}{\dot{\hat{\Phi}}}_{\rm S}^2+\frac{1}{2}C_{1}{\dot{\hat{\Phi}}}_1^2+\frac{1}{2}C_{2}{\dot{\hat{\Phi}}}_2^2+\frac{1}{2}C_{\rm L}{\dot{\hat{\Phi}}}_{\rm L}^2+E_{\rm J1}\cos\left(\frac{\hat{\Phi}_1}{\varphi_0}\right)+E_{\rm J2}\cos\left(\frac{\hat{\Phi}_2}{\varphi_0}\right)-\frac{1}{2}E_{\rm L}\left(\frac{\hat{\Phi}_{\rm L}}{\varphi_0}\right)^2,
\end{equation}
where $\varphi_0=\hbar/2e$ is the reduced flux quantum, $E_{\rm L}=\varphi_0^2/L$, and  $\hat{\Phi}_{\rm 1,2,L,S}$ are the branch flux operators. We have the following constraints due to flux quantization in the two SQUID loops, $\hat{\Phi}_{\rm L}-\hat{\Phi}_1={\Phi}_{\rm 1e}$ and $\hat{\Phi}_2-\hat{\Phi}_{\rm L}={\Phi}_{\rm 2e}$. Hence, the branch flux operators $\hat{\Phi}_1$, $\hat{\Phi}_2$, and $\hat{\Phi}_{\rm L}$ are not independent. Since the sum of all capacitances in parallel gives the total capacitance, we can assume that the shunting capacitance $C_{\rm S}$ is distributed equally to junctions $\rm J_1$ and $\rm J_2$ as additional capacitance to $C_{1,2}$. Then, using Kirchoff's voltage law, the Lagrangian can be rewritten as
\begin{equation}
    \hat{\mathcal{L}}=\frac{1}{2}\left(C_{1}+\frac{C_{\rm S}}{2}\right){\dot{\hat{\Phi}}}_1^2+\frac{1}{2}\left(C_{2}+\frac{C_{\rm S}}{2}\right){\dot{\hat{\Phi}}}_2^2+\frac{1}{2}C_{\rm L}{\dot{\hat{\Phi}}}_{\rm L}^2+E_{\rm J1}\cos\left(\frac{\hat{\Phi}_1}{\varphi_0}\right)+E_{\rm J2}\cos\left(\frac{\hat{\Phi}_2}{\varphi_0}\right)-\frac{1}{2}E_{\rm L}\left(\frac{\hat{\Phi}_{\rm L}}{\varphi_0}\right)^2,
\end{equation}

Following Ref.~\cite{you2019circuit}, given that the circuit loops threaded by time-dependent fluxes have three branches and two meshes, we a single degree of freedom in the circuit. We pick this free variable as ${\Tilde{\hat{\Phi}}}$ to govern the dynamics between the branch fluxes that are constrained by external flux loops $\Phi_{\rm 1e}$ and $\Phi_{\rm 2e}$, defined through ${\Tilde{\hat{\Phi}}}=m_{\rm L}{\hat{\Phi}}_{\rm L}+m_1{\hat{\Phi}}_1+m_2{\hat{\Phi}}_2$ where $m_{\rm L}$, $m_1$, and $m_2$ are weights that allocate the external flux to the different branches. Then we can solve for branch flux operators ${\hat{\Phi}}_{\rm L}$, ${\hat{\Phi}}_1$, and ${\hat{\Phi}}_2$ in terms of ${\Tilde{\hat{\Phi}}}$, ${\Phi}_{\rm 1e}$, and ${\Phi}_{\rm 2e}$, and obtain
\begin{align}
    {\hat{\Phi}}_{\rm L}=\frac{1}{m_\Sigma}\left[\Tilde{\hat{\Phi}}+m_1\Phi_{\rm 1e}-m_2\Phi_{\rm 2e}\right],\\
    {\hat{\Phi}}_1=\frac{1}{m_\Sigma}\left[\Tilde{\hat{\Phi}}-(m_2+m_{\rm L})\Phi_{\rm 1e}-m_2\Phi_{\rm 2e}\right],\\
    {\hat{\Phi}}_2=\frac{1}{m_\Sigma}\left[\Tilde{\hat{\Phi}}+m_{\rm 1}\Phi_{\rm 1e}+(m_1+m_{\rm L})\Phi_{\rm 2e}\right],
\end{align}
where $m_\Sigma=m_1+m_2+m_{\rm L}$. Substituting the above expressions into the Lagrangian, applying Legendre transformation, and neglecting the terms $\propto\dot{\Phi}_{\rm 1e}^2, \dot{\Phi}_{\rm 2e}^2, \dot{\Phi}_{\rm 1e}\dot{\Phi}_{\rm 2e},$ we obtain
\begin{multline}\label{eq_lagrangian_transformed}
    \hat{{\mathcal{L}}}=4E_{\rm C}\hat{n}^2+\frac{\dot{{\Tilde{\hat{\Phi}}}}}{m_\Sigma^2}
    \bigg\{\bigg[\bigg(C_2+\frac{C_{\rm S}}{2}+C_{\rm L}\bigg)m_1-\bigg(C_1+\frac{C_{\rm S}}{2}\bigg)m_2-\bigg(C_1+\frac{C_{\rm S}}{2}\bigg)m_{\rm L}\bigg]\dot{\Phi}_{\rm 1e}\\+\bigg[\bigg(C_2+\frac{C_{\rm S}}{2}\bigg)m_1-\bigg(C_1+\frac{C_{\rm S}}{2}+C_{\rm L}\bigg)m_2+\bigg(C_2+\frac{C_{\rm S}}{2}\bigg)m_{\rm L}\bigg]\dot{\Phi}_{\rm 2e}
    \bigg\}+E_{\rm J1}\cos\left(\frac{\hat{\Phi}_1}{\varphi_0}\right)+E_{\rm J2}\cos\left(\frac{\hat{\Phi}_2}{\varphi_0}\right)-\frac{1}{2}E_{\rm L}\left(\frac{\hat{\Phi}_{\rm L}}{\varphi_0}\right)^2,
\end{multline}
where the reduced charge operator $\hat{n}=in_0(\hat{a}^{\dagger}-\hat{a})$ and flux operator $\Tilde{\hat{\Phi}}=\Phi_{\rm zps}(\hat{a}^{\dagger}+\hat{a})$ are expressed in terms of the bosonic creation $(\hat a^\dagger)$ and annihilation operators $(\hat a)$, and $\Phi_{\rm zps}$ is the zero point flux fluctuations. Moreover, $E_{\rm C}=e^2/2C_\Sigma$, where $C_\Sigma=C_1+C_2+C_{\rm L}+C_{\rm S}$. There are terms that mix ${\dot{\Tilde{\Phi}}}$ with the external fluxes which can be eliminated by choosing $m_{\rm L}$, $m_1$, and $m_2$ such that the second term on the right hand side of Eq.~(\ref{eq_lagrangian_transformed}) is 0. This gives us the conditions $(C_2+C_{\rm S}/2+C_{\rm L})m_1-(C_1+C_{\rm S}/2)m_2-(C_1+C_{\rm S}/2)m_{\rm L}=0$ and $(C_2+C_{\rm S}/2)m_1-(C_1+C_{\rm S}/2+C_{\rm L})m_2-(C_2+C_{\rm S}/2)m_{\rm L}=0$. Since we have 2 equations and 3 unknowns, we have one degree of freedom. Using $m_2$ as the free variable, the solutions to the set of equations above are $m_{\rm L}=m_2 C_{\rm L}/(C_2+C_{\rm S}/2)$, $m_1=m_2 (C_1+C_{\rm S}/2)/(C_2+C_{\rm S}/2)$, $m_2=m_2$, and $m_\Sigma=m_2 C_\Sigma/(C_2+C_{\rm S}/2)$. We can simplify further by setting $m_2=(C_2+C_{\rm S}/2)/C_\Sigma$ since it is the free variable, and it allows us to make the previous arbitrary choice of $m_\Sigma=1$. Furthermore, considering $C_{\rm L}=0$, assuming symmetric junctions where $E_{\rm J2}=E_{\rm J1}=E_{\rm J}$,  and using the trigonometric identity $\cos A+\cos B=2\cos\left((A+B)/2\right)\cos\left((A-B)/2\right)$, we obtain 
\begin{multline}
    \hat{\mathcal{L}}=4E_{\rm C}\hat{n}^2+\frac{1}{2}E_{\rm L}\bigg(\frac{\Tilde{\hat{\Phi}}}{\varphi_0}-\frac{C_2+C_{\rm S}/2}{C_\Sigma}\frac{\Phi_{\rm 2e}}{\varphi_0}+\frac{C_1+C_{\rm S}/2}{C_\Sigma}\frac{\Phi_{\rm 1e}}{\varphi_0}\bigg)^2\\+2E_{\rm J}\cos\left(\frac{{\Phi}_{\rm 1e}+{\Phi}_{\rm 2e}}{2\varphi_0}\right)\cos\left(\frac{\Tilde{\hat{\Phi}}}{\varphi_0}+\frac{C_1-C_2}{2C_\Sigma}\frac{{\Phi}_{\rm 1e}+{\Phi}_{\rm 2e}}{\varphi_0}\right).
\end{multline}
We take $C_1=C_2$. Then we can define the Hamiltonian in the lab frame as
\begin{equation}
    \hat{H}_{\rm lab}=4E_{\rm C}\hat{n}^2+\frac{1}{2}E_{\rm L}\left(\frac{{\Tilde{\hat{\Phi}}}+\Phi_{\Delta}}{\varphi_0}\right)^2-2E_{\rm J}\cos\left(\frac{\Phi_\Sigma}{\varphi_0}\right)\space\cos\left(\frac{\Tilde{\hat{\Phi}}}{\varphi_0}\right),
\end{equation}
where $\Phi_{\Delta}=(\Phi_{\rm 1e}-\Phi_{\rm 2e})/2$ and $\Phi_{\Sigma}=(\Phi_{\rm 1e}+\Phi_{\rm 2e})/2$. For STS design, $\Phi_{\rm 1e}=\Phi_{\rm 2e}=\Phi_{\rm e}$, and $\Phi_\Delta=0$ so that we can express the Hamiltonian as
\begin{equation}
    \hat{H}_{\rm lab}=4E_{\rm C}\hat{n}^2+\frac{1}{2}E_{\rm L}\left(\frac{{\Tilde{\hat{\Phi}}}}{\varphi_0}\right)^2-2E_{\rm J}\cos\left(\frac{\Phi_{\rm e}}{\varphi_0}\right)\space\cos\left(\frac{\Tilde{\hat{\Phi}}}{\varphi_0}\right),
    \label{eq:STS1}
\end{equation}
where the external flux $\Phi_{\rm e}/\varphi_0=F+\delta f{\rm cos}(\omega_{\rm p}t)$ depends on the static flux $F$ and oscillating pump flux. Using our previous definitions for the reduced charge and flux operators in terms of bosonic operators, and expanding the cosine potential, we obtain
\begin{equation}
    \hat{H}_{\rm lab}=-4E_{\rm C} n_0^2(\hat{a}^\dagger-\hat{a})^2+\frac{1}{2}E_{\rm L}\left(\frac{\Phi_{\rm zps}}{\varphi_0}\right)^2(\hat{a}^\dagger+\hat{a})^2-2E_{\rm J}\cos\left(\frac{\Phi_{\rm e}}{\varphi_0}\right)\sum_n\frac{(-1)^n}{(2n)!}\left[\frac{\Phi_{\rm zps}}{\varphi_0}(\hat{a}^\dagger+\hat{a})\right]^{2n}.
\end{equation}
Following Ref.~\cite{bhandari2024symmetrically}, for weak modulation depth $\delta f\approx\Phi_{\rm zps}/\varphi_0$, we will keep only up to $\mathcal{O}(\Phi_{\rm zps}^4)$ contributions in the Hamiltonian and expand the cosine potential to fourth order. Finally, going to rotating frame defined through $\hat{a}\rightarrow\hat{a}e^{-i\omega_{\rm p} t/2}$, the Hamiltonian reduces to
\begin{multline}
    \hat{H}=-4E_{\rm C} n_0^2(\hat{a}^\dagger e^{i\omega_{\rm p} t/2}-\hat{a}e^{-i\omega_{\rm p} t/2})^2+\left(\frac{E_{\rm L}}{2}+E_{\rm J}\cos\left(\frac{\Phi_{\rm e}}{\varphi_0}\right)\right)\left(\frac{\Phi_{\rm zps}}{\varphi_0}\right)^2\left[\hat{a}^\dagger e^{i\omega_{\rm p} t/2}+\hat{a}e^{-i\omega_{\rm p} t/2}\right]^2\\
    -\frac{E_{\rm J}}{12}\cos\left(\frac{\Phi_{\rm e}}{\varphi_0}\right)\left(\frac{\Phi_{\rm zps}}{\varphi_0}\right)^4\left[\hat{a}^\dagger e^{i\omega_{\rm p} t/2}+\hat{a}e^{-i\omega_{\rm p} t/2}\right]^4.
\end{multline}
Using Jacobi-Anger expansion for $\cos(\Phi_{\rm e}/\varphi_0)$~\cite{boutin2017effect}, and ignoring the higher orders of the Bessel function of $\delta f\ll1$, following Ref.~\cite{boutin2017effect}, we can rewrite the Hamiltonian as
\begin{multline}
    \hat{H}=-4E_{\rm C} n_0^2(\hat{a}^\dagger e^{i\omega_{\rm p} t/2}-\hat{a}e^{-i\omega_{\rm p} t/2})^2+\left(\frac{\Phi_{\rm zps}}{\varphi_0}\right)^2\bigg(\frac{E_{\rm L}}{2}+E_{\rm J}^{(0)}+E_{\rm J}^{(1)}\cos(\omega_{\rm p} t)+E_{\rm J}^{(2)}\cos(2\omega_{\rm p} t)\bigg)\left[\hat{a}^\dagger e^{i\omega_{\rm p} t/2}+\hat{a}e^{-i\omega_{\rm p} t/2}\right]^2\\
    -\frac{1}{12}\left(\frac{\Phi_{\rm zps}}{\varphi_0}\right)^4(E_{\rm J}^{(0)}+E_{\rm J}^{(1)}\cos(\omega_{\rm p} t)+E_{\rm J}^{(2)}\cos(2\omega_{\rm p} t))\left[\hat{a}^\dagger e^{i\omega_{\rm p} t/2}+\hat{a}e^{-i\omega_{\rm p} t/2}\right]^4,
\end{multline}
where, following Ref.~\cite{boutin2017effect}, the Fourier coefficients are defined as $E_{\rm J}^{(0)}=E_{\rm J}J_0(\delta f)\cos F$, $E_{\rm J}^{(1)}=-2E_{\rm J}J_{1}(\delta f)\sin F$, $E_{\rm J}^{(2)}=-2E_{\rm J}J_2(\delta f)\cos F$, where $J_{m}$ is the Bessel function of the $m$th order. For small amplitude of flux pump ($\delta f\ll1$), these coefficients can be expressed as $E_{\rm J}^{(0)}\approx E_{\rm J}\cos F$, $E_{\rm J}^{(1)}\approx-E_{\rm J}\delta f\sin F$, and $E_{\rm J}^{(2)}=-(E_{\rm J}\delta f^2\cos F)/4$~\cite{boutin2017effect}. Ignoring the fast oscillating terms thorugh RWA, we obtain the time-independent Hamiltonian
\begin{equation}
    \hat{H}_{\rm STS}=\Delta a^\dagger a+\frac{\lambda}{2}(\hat{a}^{\dagger^2}+\hat{a}^2)+K\hat{a}^{\dagger^2}\hat{a}^2+\Lambda(\hat{a}^{\dagger^3}\hat{a}+\hat{a}^\dagger \hat{a}^3)+\zeta(\hat{a}^{\dagger^4}+\hat{a}^4),
    \label{eq:HSTS beforefinal}
\end{equation}
where $\Delta=\omega_{\rm a}-\omega_{\rm p}/2+2K$, $\omega_{\rm a}=\sqrt{8E_{\rm C}(E_{\rm L}+2E_{\rm J}\cos F)}$, $\lambda=E_{\rm J}^{(1)}(\Phi_{\rm zps}/\varphi_0)^2$, $K=-E_{\rm J}^{(0)}(\Phi_{\rm zps}/\varphi_0)^4/2$, $\Lambda=-E_{\rm J}^{(1)}(\Phi_{\rm zps}/\varphi_0)^4/6$, and $\zeta=-E_{\rm J}^{(2)}(\Phi_{\rm zps}/{\varphi_0})^4/24$. In our simulations for the STS performance in the main text, we use $C_{\Sigma}=4\space {\rm pF}$, $L_{\rm J}=80\space{\rm pH}$, and $L=100\space{\rm pH}$. While higher order Bessel functions and zero point flux fluctuations can be included in this analysis, based on these parameters and for modulation depth $\delta f\approx\Phi_{\rm zps}/\varphi_0$, they are negligible. Similar to the SQUID case, Kerr nonlinearity is proportional to the $\cos F$, whereas the pumping strength $\lambda$ and the extra nonideality $\Lambda$ are proportional to $\sin F$. Just like flux-pumped JPA, we need to tune the static flux at $F=-\pi/2$ to cancel Kerr nonlinearity. Based on Eq.~(\ref{eq:STS1}), tuning the static flux at $F=-\pi/2$ will leave us with an LC oscillator, which is a linear resonator we can use for signal amplification. Therefore, unlike symmetric DC SQUID, we can operate STS at static flux $F=-\pi/2$ and cancel Kerr nonlinearity entirely while keeping the pumping term. Hence, we can write the static effective Hamiltonian for our STS design as
\begin{equation}
    \hat{H}_{\rm STS}=\Delta a^\dagger a+\frac{\lambda}{2}(\hat{a}^{\dagger^2}+\hat{a}^2)+\Lambda(\hat{a}^{\dagger^3}\hat{a}+\hat{a}^\dagger \hat{a}^3),
    \label{eq:HSTS final}
\end{equation}
where $\Delta=\omega_{\rm a}-\omega_{\rm p}/2$, $\omega_{\rm a}=\sqrt{8E_{\rm C}E_{\rm L}}$. The first two terms of the Hamiltonian correspond to the quantum-limited degenerate parametric amplifier (DPA), whereas the last term is the nonideality that deviates our amplifier from the quantum limit. However, the nonideality $\Lambda$ is much smaller in amplitude than Kerr nonlinearity since $|\Lambda/K|=\delta f \tan F\ll1$ for any static flux $F$. It is also much smaller than the pump strength in amplitude since $|\Lambda/\lambda|\propto(\Phi_{\rm zps}/\varphi_0)^2\ll1$. Therefore, Kerr nonlinearity is the dominant nonideality that deviates the amplifier more from the quantum limit compared to $\Lambda$. Since STS with LC branch allows us to cancel $K$, our amplifier effectively operates at the quantum limit and is an improvement over the flux-pumped JPA's. In the case where the two SQUID loops are threaded asymmetrically where $\Phi_{\rm 1e}\neq\Phi_{\rm 2e}$ and $E_{\rm J1}\neq E_{\rm J2}$, the circuit Hamiltonian will also contain odd orders of $\Phi_{\rm zps}$. However, these odd orders vanish under RWA~\cite{bhandari2024symmetrically} and the form of the STS static effective Hamiltonian is maintained with redefined parameters, meaning that Kerr-free operation is still possible.

\subsection{STS Design with Josephson Junctions}
\label{app:STSquant}

\begin{figure}[h!]
    \centering
    \includegraphics[width=0.2\linewidth]{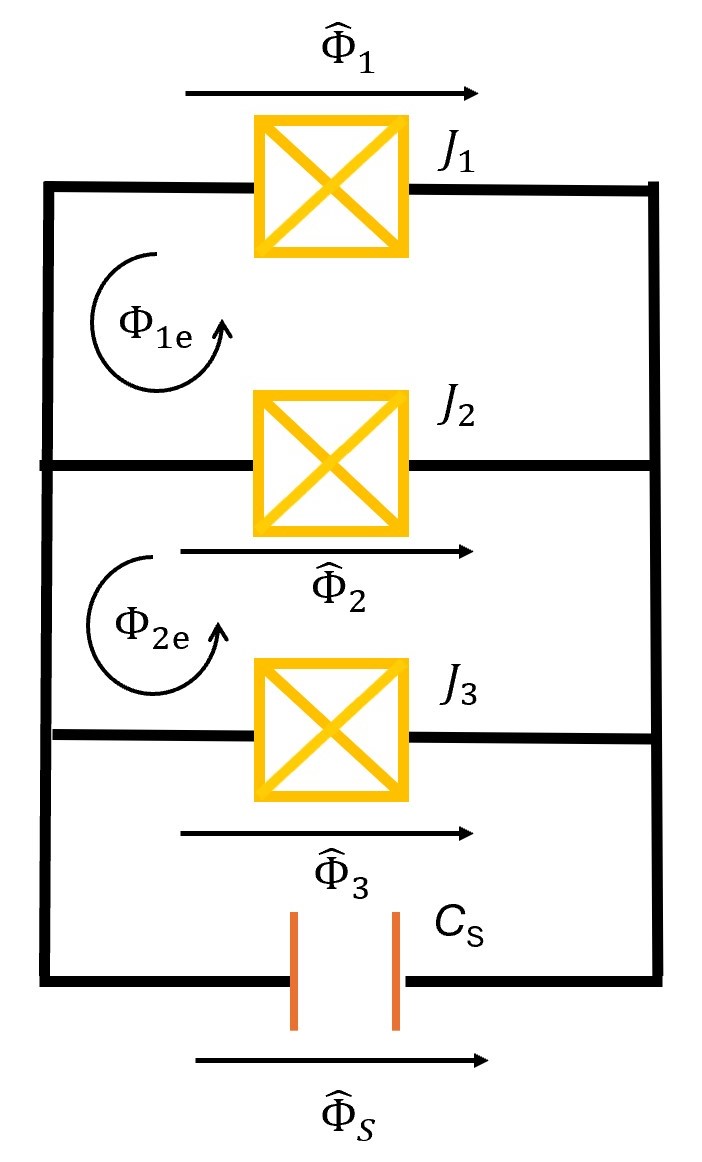}
    \caption{STS design~\cite{bhandari2024symmetrically} shunted by capacitance $C_{\rm S}$. The junctions ${\rm J}_{1}$ and ${\rm J}_{3}$ compose the ``SQUID branch" whereas the junction ${\rm J}_{2}$ gives the ``transmon branch". The loop between ${\rm J}_{1}$ and ${\rm J}_{2}$ is threaded by an external flux $\Phi_{\rm 1e}$ whereas the loop between ${\rm J}_{2}$ and ${\rm J}_{3}$ is threaded by an external flux $\Phi_{\rm 2e}$. $\hat{\Phi}_{\rm 1,2,3,S}$ are the branch flux operators.}
    \label{fig:STS}
\end{figure}

We now study the STS design discussed in Ref.~\cite{bhandari2024symmetrically} where the transmon branch consists of a Josephson junction ${\rm J}_2$ with capacitance $C_2$ as shown in Fig.~\ref{fig:STS}. Assuming $E_{\rm J1}=E_{\rm J2}=E_{\rm J3}=E_{\rm J}$, $\Phi_{\rm 1e}=\Phi_{\rm 2e}=\Phi_{\rm e}$, and following Ref.~\cite{bhandari2024symmetrically}, we can rewrite the Hamiltonian in Eq.~(\ref{eq:STS1}) as
\begin{equation}
    \hat{\cal H}_{\rm lab,STS}=4E_{\rm C}\hat{n}^2-E_{\rm J} \cos\left(\frac{\Tilde{\hat{\Phi}}}{\varphi_0}\right)-2E_{\rm J}\cos\left(\frac{\Phi_{\rm e}}{\varphi_0}\right)\cos\left(\frac{\Tilde{\hat{\Phi}}}{\varphi_0}\right),
    \label{eq:STSreal1}
\end{equation}
where $\varphi_0=\hbar/2e$ is the reduced flux quantum, ${\hat{n}}=in_0(\hat{a}^\dagger-\hat{a})$ and $\Tilde{\hat{\Phi}} = \Phi_{\rm zps}(\hat{a}^\dagger+\hat{a})$, and $\Phi_{\rm zps}$ is the zero point phase spread around the Josephson junction. Following the derivations in Ref.~\cite{bhandari2024symmetrically}, the static effective STS Hamiltonian can be written in the same form as the Kerr-free STS amplifier
\begin{equation}
    \hat{\cal H}_{\rm STS}=\Delta a^\dagger a+\frac{\lambda}{2}(\hat{a}^{\dagger^2}+\hat{a}^2)+K\hat{a}^{\dagger^2}\hat{a}^2+\Lambda(\hat{a}^{\dagger^3}\hat{a}+\hat{a}^\dagger \hat{a}^3),
\end{equation}
where the detailed definition of each coefficient can be found in Ref.~\cite{bhandari2024symmetrically}. However, unlike our STS design with the LC branch, the transmon branch in this design provides Kerr nonlinearity $K=-E_{\rm C}/2$ which cannot be cancelled by static flux tuning. Therefore, having an LC branch is crucial for obtaining a Kerr-free amplifier. 

\section{Harmonic Balance and Stability Diagram}
\label{app:stability}
We analyze the Quantum Langevin equation (QLE) for mode $\hat{a}$ derived from the input-output theory~\cite{walls2008quantum} 
\begin{equation}
    \dot{\hat{a}}=i[\hat{H}_{\rm STS},\hat{a}]-\frac{\kappa}{2}\hat{a}+\sqrt{\kappa}\hat{a}_{\rm in},
    \label{eq:QLE}
\end{equation}
where $\kappa$ is the cavity decay rate and the static effective Hamiltonian $\hat{H}_{\rm STS}$ is defined in the main text. We treat the input and output fields as classical drives $\alpha_{\rm in(out)}=\sum_{x}\alpha_xe^{-i\omega_x t}$ where $\omega_x=m\omega_p+n\omega_s$ represents all possible harmonics, with $m,n$ being integer values. To solve the QLE, we apply the semiclassical harmonic balance solution
\begin{equation}
\alpha(t)=\bra{\alpha}\hat{a}(t)\ket{\alpha}=\sum_{\omega=-\omega_x}^{\omega_x}\alpha(\omega)e^{-i\omega t},
\end{equation}
which runs over all the harmonics mentioned before. This can be expressed as
\begin{equation}
\alpha(t)=\sum_{\omega=0}^{\omega_x}\bigg(\alpha(\omega)e^{-i\omega t}+\alpha(-\omega)e^{i\omega t}\bigg).
\label{eq:harbal}
\end{equation}
We can express the second term in the sum as a positive frequency Fourier component. For that, we first look into the relation between the flux and charge operators $\dot{\hat{\Phi}}=\hat{Q}/C_{\rm c}$, where $\hat{\Phi}=\Phi_{\rm zps}(\hat{a}+\hat{a}^{\dagger})$, $\hat{Q}=iq_{\rm zps}(\hat{a}^{\dagger}-\hat{a})$, and $C_{\rm c}$ is the capacitance of our lumped-element amplifier circuit. Based on the Fourier transform
\begin{equation}
    \hat{a}(t)=\frac{1}{\sqrt{2\pi}}\int_{-\infty}^{\infty}\overline{a}[\omega]e^{-i\omega t}d\omega,
    \label{eq:FT}
\end{equation}
\begin{equation}
    \overline{\hat{a}}[\omega]=\frac{1}{\sqrt{2\pi}}\int_{-\infty}^{\infty}\hat{a}(t)e^{i\omega t}dt,
\end{equation}
where $\overline{a}[\omega]$ is the Fourier transform of the annihilation operator. We can then say that $\hat{\dot{\Phi}}=-i\omega\hat{\Phi}$. Using the relation between the flux and charge operators discussed above, we express the relation between the creation and annihilation operator as
\begin{equation}
    \hat{a}^{\dagger}(t)=\frac{\omega_{\rm c}-\omega}{\omega_{\rm c}+\omega}\hat{a}(t),
\end{equation}
where $\omega_{\rm c}=q_{\rm zps}/(C_{\rm c}\Phi_{\rm zps})$ is the cavity resonance frequency. Then the relation between the negative frequency Fourier component of the annihilation operator and the positive frequency Fourier component of the creation operator is given by
\begin{equation}
    \overline{\hat{a}}^{\dagger}[-\omega]=\frac{\omega_{\rm c}-\omega}{\omega_{\rm c}+\omega}\overline{\hat{a}}[\omega].
\end{equation}
Taking the Hermitian conjugate of this expression, we get
\begin{equation}
    \overline{\hat{a}}[-\omega]=\frac{\omega_{\rm c}-\omega}{\omega_{\rm c}+\omega}\overline{\hat{a}}^{\dagger}[\omega].
\end{equation}
Then we can rewrite second term inside the sum in Eq.~(\ref{eq:harbal}), and get

\begin{equation}
\alpha(t)=\sum_{\omega=0}^{\omega_x}\bigg(\alpha(\omega)e^{-i\omega t}+\frac{\omega_{\rm c}-\omega_x}{\omega_{\rm c}+\omega_x}\alpha^*(\omega)e^{i\omega t}\bigg).
\end{equation}
For the rest of the paper, we will use the notation $\alpha_{x}=\alpha(\omega_x)$. We neglect the out-of-band harmonics, and only take into account the harmonics at frequencies $\omega_{\rm s}, \omega_{\rm i}$, and $\omega_{\rm p}/2$. Since we assume that these harmonics are in resonance with the cavity frequency, the second term of the harmonic balance solution vanishes. The equation of motion in Eq.~(\ref{eq:QLE}) is reduced to three equations for the amplitudes of these harmonics ($\alpha_{\rm s}, \alpha_{\rm i}$, and $\alpha_{\rm h}$ respectively) as follows:

\begin{align}
    (\omega-\Delta+i\frac{\kappa}{2})\alpha_{\rm s}=i\sqrt{\kappa}\alpha_{\rm in,s}+\lambda\alpha_{\rm i}^*+3\Lambda[(|\alpha_{\rm i}|^2+2|\alpha_{\rm s}|^2+2|\alpha_{\rm h}|^2)\alpha_{\rm i}^*+(\alpha_{\rm h}^2+\alpha_{\rm h}^{*2}+\alpha_{\rm i}\alpha_{\rm s})\alpha_{\rm s}], \label{eq:as}\\
    (-\omega-\Delta+i\frac{\kappa}{2})\alpha_{\rm i}=i\sqrt{\kappa}\alpha_{\rm in,i}+\lambda\alpha_{\rm s}^*+3\Lambda[(|\alpha_{\rm s}|^2+2|\alpha_{\rm i}|^2+2|\alpha_{\rm h}|^2)\alpha_{\rm s}^*+(\alpha_{\rm h}^2+\alpha_{\rm h}^{*2}+\alpha_{\rm i}\alpha_{\rm s})\alpha_{\rm i}], \label{eq:ai}\\
    (-\Delta+i\frac{\kappa}{2})\alpha_{\rm h}=i\sqrt{\kappa}\alpha_{\rm in,h}+\lambda\alpha_{\rm h}^*+3\Lambda[(|\alpha_{\rm h}|^2+2|\alpha_{\rm s}|^2+2|\alpha_{\rm i}|^2)\alpha_{\rm h}^*+[2(\alpha_{\rm i}\alpha_{\rm s}+\alpha_{\rm i}^*\alpha_{\rm s}^*)+\frac{1}{3}\alpha_{\rm h}^2]\alpha_{\rm h}],
    \label{eq:ah}
\end{align}

where $\omega=\omega_{\rm s}-\omega_{\rm p}/2$ is the signal detuning, and $\alpha_{\rm in,s},\alpha_{\rm in,i}$, and $\alpha_{\rm in,h}$ are the classical drive strengths at corresponding harmonics. To determine the stability regions of our amplifier, we solve for the cavity amplitude $\alpha_{\rm h}$ in Eq.~(\ref{eq:ah}). We assume small-signal approximation where the terms $|\alpha_{\rm s}|^2, |\alpha_{\rm i}|^2, \alpha_{\rm i}\alpha_{\rm s}$, and $\alpha_{\rm i}^*\alpha_{\rm s}^*$ can be neglected. Since there is no input drive at frequency $\omega_{\rm p}/2$, we take $\alpha_{\rm in,h}=0$. Then Eq.~(\ref{eq:ah}) is simplified as
\begin{equation}
    (-\Delta+i\frac{\kappa}{2}-\Lambda\alpha_{\rm h}^2)\alpha_{\rm h}=(\lambda+3\Lambda|\alpha_{\rm h}|^2)\alpha_{\rm h}^*.
    \label{eq:simpcaveq}
\end{equation}
We can express $\alpha_{\rm h}$ as $\alpha_{\rm h}=x+iy$, where $x={\rm Re}(\alpha_{\rm h})$ and $y={\rm Im}(\alpha_{\rm h})$. Substituting this into Eq.~(\ref{eq:simpcaveq}), we can write the real part of the whole expression as
\begin{equation}
    4\Lambda x^3+(\lambda+\Delta)x+\frac{\kappa}{2}y=0,
\end{equation}
and the imaginary part as
\begin{equation}
    4\Lambda y^3+(\lambda-\Delta)y+\frac{\kappa}{2}x=0.
\end{equation}
We can see that in the limit where $\Lambda=0$, we obtain a system of linear equations with solution $(x,y)=(0,0)$, which corresponds to the case of a DPA.

To solve the system of nonlinear equations in general, we can uncouple the equations into two ninth-order polynomials. This results several different possible pairs of solutions depending on the values of $\lambda$ and $\Delta$: one solution, three solutions, five solutions, seven solutions, and nine solutions. We note that for all $\lambda$ and $\Delta$ values, the trivial solution $(x,y)=(0,0)$ is always a solution. We also further note that if $(x,y)=(r_1,r_2)$ is a non-trivial solution to the coupled equations, then $(x,y)=(-r_1,-r_2)$ is also a solution. Since the number of stable points is equivalent to the number of unique $|\alpha_{\rm h}|^2$ solutions, we can conclude that $(r_1,r_2)$ and $(-r_1,-r_2)$ give the same $|\alpha_{\rm h}|^2$ solution, since $|\alpha_{\rm h}|^2=x^2+y^2=r_1^2+r_2^2=(-r_1)^2+(-r_2)^2$. This leads us to conclude that the number of possible stable $|\alpha_{\rm h}|^2$ values is between one and five depending on the values of $\lambda$ and $\Delta$. The number of stable points can be plotted in a stability diagram as a function of $\lambda$ and $\Delta$.

\begin{figure}[h]
    \centering
    \includegraphics[width=0.8\linewidth]{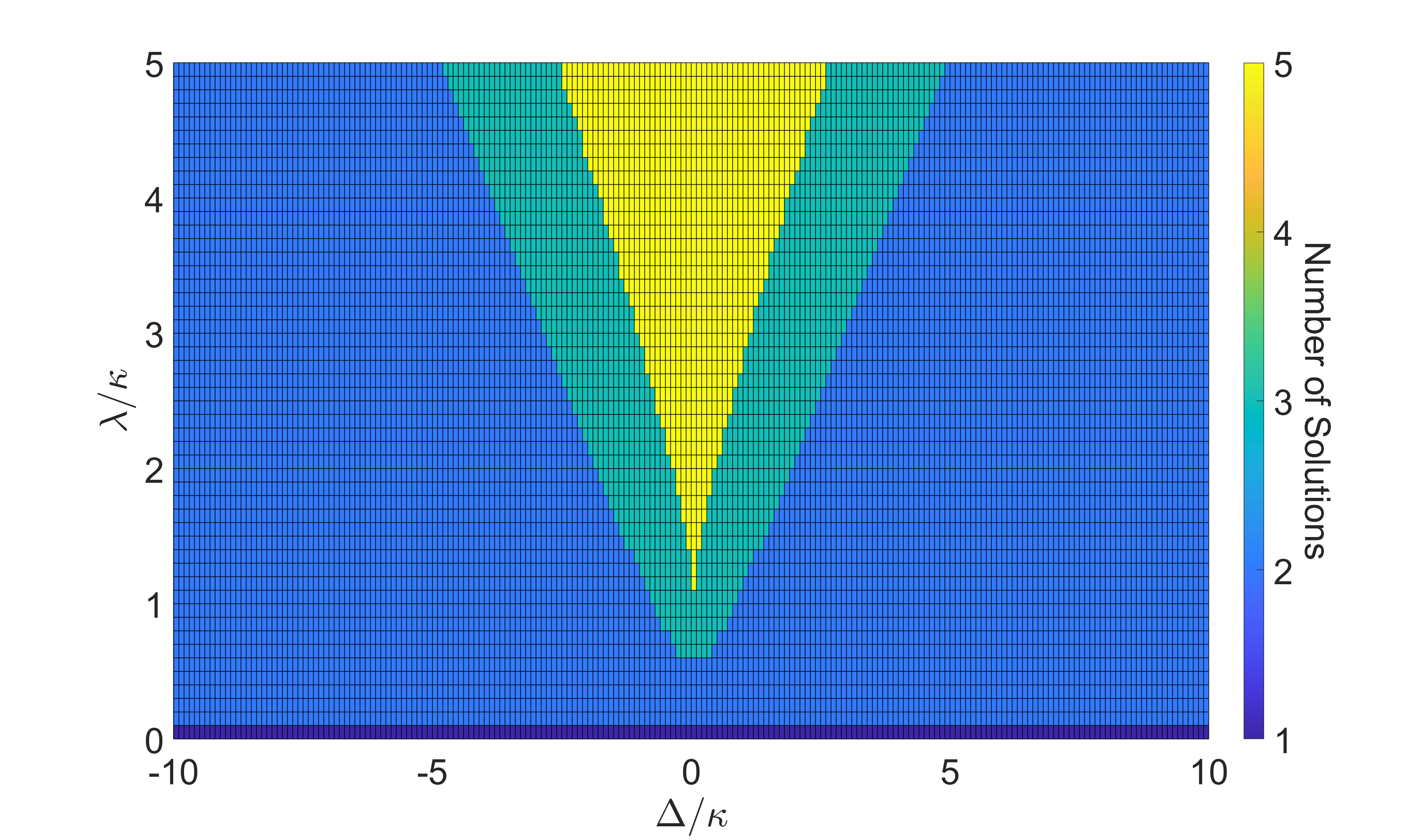}
    \caption{ The stability diagram of the Kerr-free STS amplifier for various drive and detuning values. The case of four stable $|\alpha_{\rm h}|^2$ values occur right at the border of the tri-stable and penta-stable regions.}
    \label{fig:SDP1}
\end{figure}
We find that the transition between the bi-stable and tri-stable region is given by the condition $\lambda^2-\Delta^2=\kappa^2/4$.

\begin{figure}[h]
    \centering
    \includegraphics[width=0.4\linewidth]{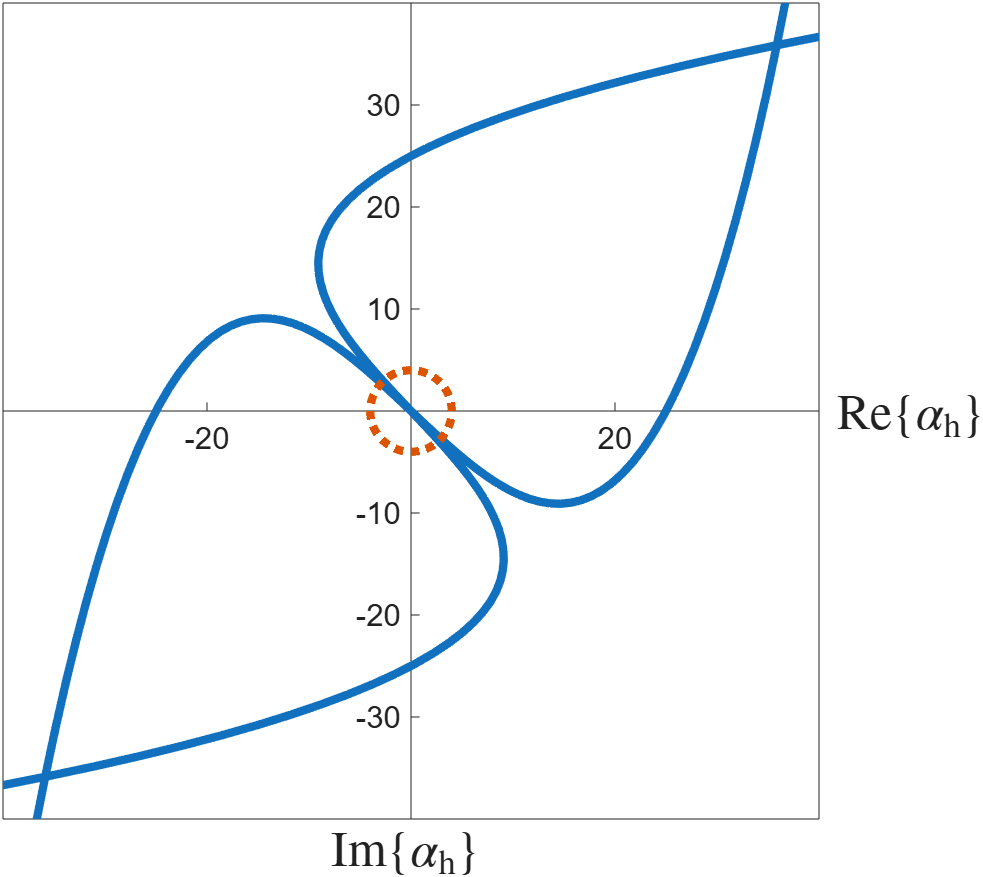}
    \caption{The stable points of $\alpha_h$ for $\Delta=0$ and $|\lambda|=0.472631\kappa$.  The intersection of the blue plots are the solutions to the coupled equations and $|\alpha_{\rm h}|^2$ is the distance from the origin to the intersection point. This corresponds to the bistable regime, since there are two unique $|\alpha_{\rm h}|^2$ values. The dashed circle is the typical total photon number in the Kerr-free STS amplifier, which is far away from the nontrivial intersection points. Therefore, the STS amplifier can be treated as an effective mono-stable system.}
    \label{fig:AhSP1}
\end{figure}

For the region where $\lambda/\kappa<0.5$, there is only bistability. Despite that, we can still treat the STS amplifier as effectively mono-stable because of how far apart the two stable points are, as shown in Figure \ref{fig:AhSP1}. Since amplifiers usually only operate at the trivial point as in the case of the DPA, as long as the photon number is small we do not have to be concerned with the other stable point coming into play.

\section{Analytical Gain Expression}
\label{app:gain}
To derive the phase-preserving gain analytically, we evaluate the coupled equations of the single-SQUID JPA and then obtain the results for STS amplifier for $K=0$. Under the assumption that we can operate an amplifier at $\alpha_{\rm h}=0$, as discussed in previous section, we can rewrite these equations as

\begin{align}
    \left(\omega-(\Delta+2K(2|\alpha_{\rm i}|^2+|\alpha_{\rm s}|^2)+3\Lambda\alpha_{\rm i}\alpha_{\rm s})+i\frac{\kappa}{2}\right)\alpha_{\rm s}=i\sqrt{\kappa}\alpha_{\rm in,s}+\left(\lambda+3\Lambda(|\alpha_{\rm i}|^2+2|\alpha_{\rm s}|^2)\right)\alpha_{\rm i}^*, \label{eq:as_g}\\
    \left(-\omega-(\Delta+2K(|\alpha_{\rm i}|^2+2|\alpha_{\rm s}|^2)+3\Lambda\alpha_{\rm i}\alpha_{\rm s})+i\frac{\kappa}{2}\right)\alpha_{\rm i}=i\sqrt{\kappa}\alpha_{\rm in,i}+\left(\lambda+3\Lambda(|\alpha_{\rm s}|^2+2|\alpha_{\rm i}|^2)\right)\alpha_{\rm s}^*\label{eq:ai_g}.
\end{align}
For large gain ($G\gg1$), we can assume that $n_{\rm s}\approx n_{\rm i}$ where $n_{\rm s,i}=|\alpha_{\rm s,i}|^2$~\cite{PhysRevApplied10054020}. Then, we can rewrite these equations as
\begin{align}
    \left(\omega-\Delta_{\rm eff}+i\frac{\kappa}{2}\right)\alpha_{\rm s}=i\sqrt{\kappa}\alpha_{\rm in,s}+\lambda_{\rm eff}\alpha_{\rm i}^*, \label{eq:as_g}\\
    \left(-\omega-\Delta_{\rm eff}^*-i\frac{\kappa}{2}\right)\alpha_{\rm i}^*=-i\sqrt{\kappa}\alpha_{\rm in,i}^*+\lambda_{\rm eff}^*\alpha_{\rm s},\label{eq:ai_g}
\end{align}
where $\Delta_{\rm eff}=\Delta+6Kn_{\rm s}+3\Lambda\alpha_{\rm i}\alpha_{\rm s}$, and $\lambda_{\rm eff}=\lambda+9\Lambda n_{\rm s}$, and we have taken the complex conjugate of the equation for the idler harmonic. The effective detuning and squeezing drive contain corrections from the nonidealities $K$ and $\Lambda$. Let us express the coefficients with letters $\rm A$ and $\rm B$ such that
\begin{align}
    {\rm A}\alpha_{\rm s}-\lambda_{\rm eff}\alpha_{\rm i}^*=i\sqrt{\kappa}\alpha_{\rm in,s}, \label{eq:as_abc}\\
    {\rm B}\alpha_{\rm i}^*-\lambda_{\rm eff}^*\alpha_{\rm s}=-i\sqrt{\kappa}\alpha_{\rm in,i}^*. \label{eq:ai_abc}
\end{align}
Then we can write $\alpha_{\rm i}^*$ in terms of $\alpha_{\rm s}$, and vice versa, and then express the matrix equation between the signal and idler amplitudes and input field amplitudes
\begin{equation}
   \begin{pmatrix}
\alpha_{\rm s} \\
\alpha_{\rm i}^* 
\end{pmatrix}=i\sqrt{\kappa}
   \begin{pmatrix}
1/\rm D_1 & \rm \lambda_{\rm eff}/(BD_1)\\
\rm \lambda_{\rm eff}^*/(AD_2) & -1/\rm D_2
\end{pmatrix} 
\begin{pmatrix}
\alpha_{\rm in,s} \\
\alpha_{\rm in,i}^* 
\end{pmatrix},
\end{equation}
where $\rm D_1=A-|\lambda_{\rm eff}|^2/B$, and $\rm D_2=B-|\lambda_{\rm eff}|^2/A$. For each harmonics, we use the input output relation~\cite{walls2008quantum}
\begin{equation}
    \alpha_{\rm out}=\sqrt{\kappa}\alpha-\alpha_{\rm in}.
    \label{eq:in-out}
\end{equation}
Then, we can write the matrix for the input and output fields as
\begin{equation}
\begin{pmatrix}
\alpha_{\rm out,s} \\
\alpha_{\rm out,i}^{*} 
\end{pmatrix}=i\kappa
   \begin{pmatrix}
1/\rm D_1 & \rm \lambda_{\rm eff}/(BD_1)\\
\rm \lambda_{\rm eff}^*/(AD_2) & -1/\rm D_2
\end{pmatrix} 
\begin{pmatrix}
\alpha_{\rm in,s} \\
\alpha_{\rm in,i}^* 
\end{pmatrix}-\begin{pmatrix}
1 & 0\\
0 & 1
\end{pmatrix} 
\begin{pmatrix}
\alpha_{\rm in,s} \\
\alpha_{\rm in,i}^* 
\end{pmatrix}.
\end{equation}
The signal gain is expressed as
\begin{equation}
    G_{\rm s}=\bigg|\frac{\alpha_{\rm out,s}}{\alpha_{\rm in,s}}\bigg|^2=\left|\frac{i\kappa\left(\omega+\Delta_{\rm eff}^*+i\frac{\kappa}{2}\right)}{\left(\omega-\Delta_{\rm eff}+i\frac{\kappa}{2}\right)\left(\omega+\Delta_{\rm eff}^*+i\frac{\kappa}{2}\right)+|\lambda_{\rm eff}|^2}-1\right|^2.
\end{equation}
This is effectively the same gain expression as DPA, as discussed in~\cite{boutin2017effect}. For the STS amplifier, we take $K=0$, which has smaller corrections to the detuning compared to single-SQUID JPA's and is therefore closer to a DPA.

\subsection{Normalizing the Added Noise}\label{app:normalization of efficiency}
For our simulations in phase-preserving quantum efficiency, we note that while the added noise number $\mathcal{A}$ is unitless and represents the photon number, the input-output relation in Eq.~(\ref{eq:in-out}) suggests that the second-moment noises are in units of frequency. We therefore need to normalize the second-order noises so that they are in units of photon numbers.

\begin{figure}[b]
    \centering
    \includegraphics[width=0.5\linewidth]{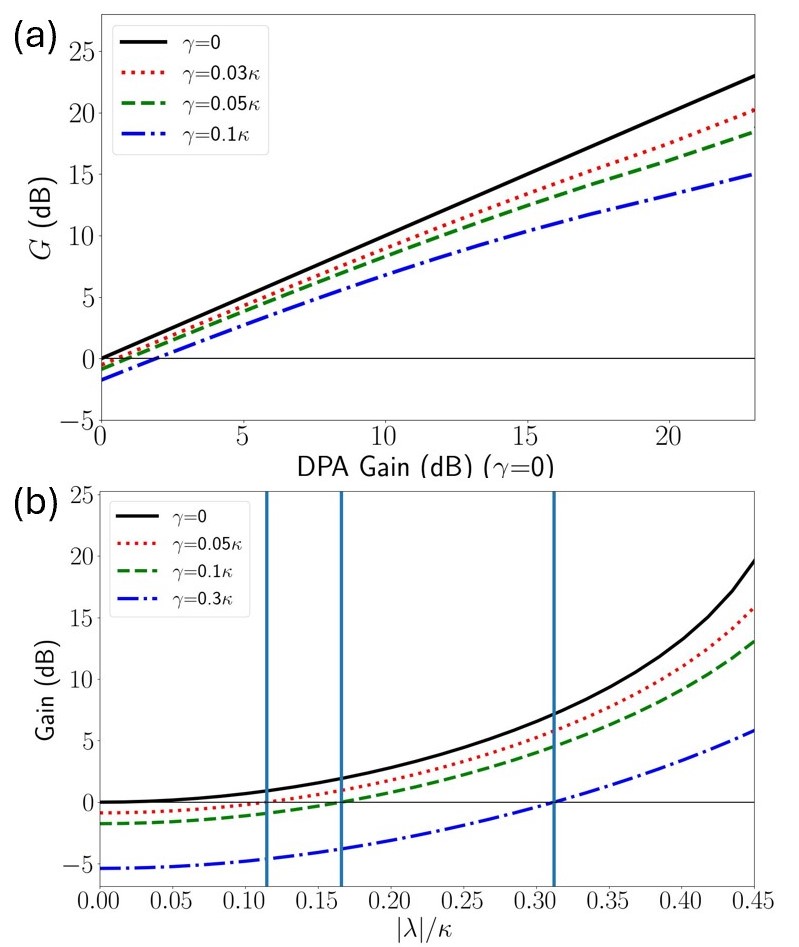}
    \caption{(a) STS amplifier phase-preserving gain against DPA gain without additional loss rate $\gamma$. (b) STS amplifier phase-preserving gain against two-photon drive strength $\lambda$ for various coupling rates $\gamma$ due to undesireable additional vacuum noise mixed to the signal. Vertical lines represent the $\lambda$ necessary to mitigate deamplification for each plot, derived from the analytical DPA gain in~\cite{boutin2017effect}. For both figures $\kappa=300\space\rm MHz$, $\Delta=\omega=0$.}
    \label{fig:extra noise}
\end{figure}

As discussed in~\cite{boutin2017effect,roy2016introduction}, for degenerate signal and idler inputs ($\omega_{\rm s}=\omega_{\rm i}$), we can write the added noise operator in terms of the input idler field $\hat{a}^{\dagger}_{\rm in}$ and write the input-output relation as
\begin{equation}
    \hat{a}_{\rm out}=g\hat{a}_{\rm in}+\sqrt{|g|^2-1}\hat{a}^{\dagger}_{\rm in},
\end{equation}
where $g$ is the signal amplitude gain, and $\omega=\omega_{\rm s}-\omega_{\rm p}/2$. Then, based on~\cite{caves2012quantum}, the added noise can be defined in terms of the idler noise
\begin{equation}
    \mathcal{A}=\frac{G-1}{G}\langle|\Delta\hat{a}_{\rm in}|^2\rangle,
\end{equation}
where $G=|g|^2$ is the phase-preserving signal gain. Based on the relation $\mathcal{A}\geq\left(1-1/G\right)/2$ derived in~\cite{caves2012quantum}, we know that a quantum-limited DPA should satisfy the lower bound of the added noise number. Hence idler noise $\langle|\Delta\hat{a}_{\rm in}|^2\rangle=1/2$ is in units of photon number. To express idler noise analytically, we first express the input field analytically, using the QLE
\begin{equation}
    \dot{\hat{a}}=i[\hat{H}_{\rm DPA}+\hat{H}_{\rm probe},\hat{a}]-\frac{\kappa}{2}\hat{a}+\sqrt{\kappa}\hat{a}_{\rm in}.
    \label{eq:QLE}
\end{equation}
Then, we express the input field in terms of the intracavity field operators in steady state
\begin{equation}
    \hat{a}_{\rm in}=\frac{1}{\sqrt{\kappa}}\left[\left(-i\omega+i\Delta+\frac{\kappa}{2}\right)\hat{a}+i\lambda\hat{a}^{\dagger}+i\epsilon_{\rm probe}\right].
\end{equation}
Since our simulations are done at center frequency $\omega=0$ where signal is resonant with the cavity, the noise in vacuum is
\begin{equation}
    \langle|\Delta\hat{a}_{\rm in}|^2\rangle=\bra{0}|\Delta\hat{a}_{\rm in}|^2\ket{0}=\frac{1}{2}\left(\frac{\kappa}{4}+\frac{|\lambda|^2}{\kappa}\right).
\end{equation}
Since we know the added noise for DPA at high gain is half a photon, we normalize the noise by a drive-dependent frequency of $(\kappa/4+|\lambda|^2/\kappa)$. When calculating the added noise for JPAs and STS, we similarly derive $\hat{a}_{\rm in}$ from QLE and calculate the noise $\langle|\Delta\hat{a}_{\rm in}|^2\rangle$ analytically. We then calculate $\mathcal{A}$ by simulating the phase-preserving gain $G$ by sweeping $\lambda$. Finally, we normalize $\mathcal{A}$ by $(\kappa/4+|\lambda_{\rm DPA}|^2/\kappa)$, where $\lambda_{\rm DPA}$ is the two-photon drive strength applied on DPA that corresponds to the gain of the given JPA or STS, which we can calculate from the analytical gain expression derived in~\cite{boutin2017effect}.

\subsection{Extra Noise}
In this paper we have assumed a lossless parametric amplifier, where the damping rate of the cavity only consists of $\kappa$ due to coupling the signal to be amplified. However, in experiments there are undesired losses in the resonator that mix additional vacuum noise to the signal~\cite{boutin2017effect}. We can treat these losses as an additional input port coupled to the cavity with an input field $\hat{b}_{\rm in}$ and coupling rate $\gamma$. The total damping rate for the STS amplifier is then $\Bar{\kappa}=\kappa+\gamma$ and the QLE is written as

\begin{equation}
    \dot{\hat{a}}=i[\hat{H}_{\rm STS},\hat{a}]-\frac{\Bar{\kappa}}{2}\hat{a}+\sqrt{\kappa}\hat{a}_{\rm in}+\sqrt{\gamma}\hat{b}_{\rm in}.
    \label{eq:QLE extra noise}
\end{equation}
In Fig.~\ref{fig:extra noise}(a), we show that a lossless ($\gamma=0$) Kerr-free STS amplifier is performing like a DPA, as seen in our previous phase-preserving gain results. However, when there is extra noise mixing with the signal due to undesired losses ($\gamma>0$), the STS amplifier deviates from DPA and gain decreases with increasing $\gamma$. We further note that extra damping due to $\gamma$ leads to deamplification as we see gain below zero. In Fig.~\ref{fig:extra noise}(b) we show how much two-photon drive strength $\lambda$ is necessary to mitigate the deamplification of the STS under various damping rates of $\gamma$. It shows that increasing damping rate $\gamma$ deamplifies the signal more and requires stronger two-photon drive to be mitigated. The zero gain threshold for the drive strengths in this figure have been calculated from the analytical gain expression for DPA. As discussed in Ref.~\cite{boutin2017effect}, the signal gain for DPA with extra loss can be derived from the input-output relation in Fourier space
\begin{equation}
    \overline{\hat{a}_{\rm out}}[\omega]=g_{S,\omega}\overline{\hat{a}_{\rm in}}[\omega]+g_{I,\omega}\overline{\hat{a}_{\rm in}}^{\dagger}[-\omega]+\sqrt{\frac{\gamma}{\kappa}}\left[(g_{S,\omega}+1)\overline{\hat{b}_{\rm in}}[\omega]+g_{I,\omega}\overline{\hat{b}_{\rm in}}^{\dagger}[-\omega]\right],
\end{equation}
where $g_{S,\omega}$ is the signal amplitude gain and $g_{I,\omega}$ is the idler amplitude gain. The DPA amplitude gain is expressed as
\begin{equation}
    g_{S,\omega}=\frac{\kappa\overline{\kappa}/2-i\kappa(\Delta+\omega)}{\Delta^2(\overline{\kappa}/2-i\omega)^2-|\lambda|^2}-1
\end{equation}

We see that the DPA gain results match exactly with the drive strengths at which STS gain is zero. This shows that our lossy STS amplifier operates at lossy DPA level performance.

\putbib[refs_sup]
\end{bibunit}
\thispagestyle{empty}
\end{document}